\documentclass[12pt]{article}

\usepackage{graphicx}
\usepackage{amsmath,amssymb} 
\usepackage{latexsym}
\usepackage{mathrsfs}
\usepackage{bbold}
\usepackage{color}
\usepackage{slashed}
\definecolor{red}{rgb}{1,0,0}
\usepackage{cancel}
\usepackage{comment}
\usepackage{cite,mathtools}
\usepackage{caption}
\usepackage{subcaption}
\definecolor{red}{rgb}{1,0,0}

\newcommand{\be}{\begin{equation}}
\newcommand{\ee}{\end{equation}}
\newcommand{\bee}{\begin{equation*}}
\newcommand{\eee}{\end{equation*}}
\newcommand{\bea}{\begin{eqnarray}}
\newcommand{\eea}{\end{eqnarray}}
\newcommand{\bean}{\begin{eqnarray*}}
\newcommand{\eean}{\end{eqnarray*}}

\def\bea{\begin{eqnarray}} \def\eea{\end{eqnarray}}
\def\be{\begin{equation}} \def\ee{\end{equation}}

\newcommand{\im}{{\operatorname{Im}}}

\newcommand{\promille}{%
  \relax\ifmmode\promillezeichen
        \else\leavevmode\(\mathsurround=0pt\promillezeichen\)\fi}
\newcommand{\promillezeichen}{%
  \kern-.05em%
  \raise.5ex\hbox{\the\scriptfont0 0}%
  \kern-.15em/\kern-.15em%
  \lower.25ex\hbox{\the\scriptfont0 00}}

\newcommand{\ir}{{_{\rm IR}}}
\newcommand{\uv}{{_{\rm UV}}}

\usepackage[colorlinks,linkcolor=black,citecolor=blue,urlcolor=blue,linktocpage]{hyperref}

\makeatletter
\def\section{\@startsection {section}{1}{\z@}{-3.5ex plus -1ex minus
 -.2ex}{2.3ex plus .2ex}{\large\bf}}
\def\subsection{\@startsection{subsection}{2}{\z@}{-3.25ex plus -1ex
minus -.2ex}{1.5ex plus .2ex}{\normalsize\bf}}
\makeatother
\makeatletter

\@addtoreset{equation}{section}

\makeatother

\begin{document}

\setcounter{page}{0}
\thispagestyle{empty}

\begin{flushright}
DESY 16-221
\end{flushright}

\vskip 8pt

\begin{center}
{\bf \LARGE {Cosmological evolution of Yukawa couplings: 
\vskip 10pt
the 5D perspective }}
\end{center}

\vskip 12pt

\begin{center}
 { \bf  Benedict von Harling$^a$ and G\'eraldine Servant$^{a,b}$}
 \end{center}

\vskip 14pt

\begin{center}
\centerline{$^{a}${\it DESY, Notkestrasse 85, 22607 Hamburg, Germany}}
\centerline{$^{b}${\it II.~Institute of Theoretical Physics, University of Hamburg, 22761 Hamburg, Germany}}

\vskip .3cm
\centerline{\tt benedict.von.harling@desy.de, geraldine.servant@desy.de}
\end{center}

\vskip 10pt

\begin{abstract}
\vskip 3pt
\noindent

The cosmological evolution of standard model Yukawa couplings may have major implications for baryogenesis.
In particular, as highlighted recently, the CKM matrix alone could be the source of $CP$-violation during electroweak baryogenesis provided that the Yukawa couplings were large and varied during the electroweak phase transition.  We provide a natural realisation of this idea in the context of Randall-Sundrum models and show that the geometrical warped approach to the fermion mass hierarchy may naturally display the desired cosmological dynamics.
The key ingredient is the coupling  of the Goldberger-Wise scalar, responsible for the IR brane stabilisation, to the bulk fermions, which modifies the fermionic profiles. This also helps alleviating the usually tight  constraints from $CP$-violation in Randall-Sundrum scenarios.
We study how the Yukawa couplings vary during the stabilisation of the Randall-Sundrum geometry and can thus induce large $CP$-violation during the electroweak phase transition. Using holography, we discuss the 4D interpretation of this dynamical interplay between flavour and electroweak symmetry breaking. 

\end{abstract}

\newpage

\tableofcontents

\vskip 13pt

\newpage

\section{Introduction}
\label{sec:intro}

The origin of the flavour structure is one of the major puzzles of the standard model (SM).
While many solutions have been proposed, the cosmological aspects of the corresponding models
 have hardly been studied. On the other hand, in many cases Yukawa couplings are dynamical and it is natural to investigate the possibility of their cosmological evolution, and whether this could help addressing open problems, like baryogenesis. 
Such questions were started to be addressed recently 
\cite{Berkooz:2004kx,Baldes:2016rqn,Baldes:2016gaf}. In particular, the CKM matrix can be the unique source of $CP$-violation for electroweak baryogenesis if Yukawa couplings vary at the same time as the Higgs is acquiring its vacuum expectation value (VEV) \cite{dynamicalyukawas}. With these motivations in mind, we are interested in studying natural realisations of Yukawa variation at the electroweak (EW) scale.
 
In this paper, our aim is to investigate the possibility of varying Yukawas in Randall-Sundrum (RS) models \cite{Randall:1999ee}.
One of the very attractive features of the RS model is that in addition to bringing a new solution to the Planck scale/weak scale hierarchy problem, it offers a new tool to understand flavour and explain the hierarchy of fermion masses \cite{Gherghetta:2000qt,Huber:2000ie,Gherghetta:2010cj}. The setup is a slice of 5D Anti-de Sitter space (AdS$_5$) which is bounded by two branes, the UV (Planck) brane where the graviton is peaked, and the IR (TeV) brane hosting  the Higgs (which therefore does not feel the UV cutoff). Fermions and gauge bosons are free to propagate in the bulk.
In this framework, the effective 4D Yukawas of SM fermions are given by the overlap of their 5D wavefunctions with the Higgs. 
Since the Higgs is localised towards the IR brane to address the Planck scale/weak scale hierarchy, 
small Yukawas are achieved if the fermions live towards the UV brane so that the overlap between the fermions and the Higgs is suppressed. On the other hand, heavy fermions such as the top quark are localised near the IR brane. This setup leads to a protection from large flavour and $CP$-violation via the so-called RS-GIM mechanism.

The key feature for flavour physics is therefore the localisation of the fermions in the AdS$_5$ slice, which determines the effective scale of  higher-dimensional  flavour-violating operators.
The profile of a fermion is  determined by its 5D  bulk mass. 
Because of the AdS$_5$ geometry, modifications of order  one in the 5D bulk mass have a substantial  impact on the fermionic profile and therefore on the effective 4D Yukawa coupling. In fact, the 4D Yukawa couplings depend exponentially on the bulk mass parameter.
Randall-Sundrum models are holographic duals of 4D strongly coupled theories. In this picture, the Higgs is part of the composite sector. 
The size of the Yukawa couplings is then determined by the degree of compositeness of the states that are identified with the SM fermions.
Indeed fermions localised near the UV brane are dual to mainly elementary states leading to small Yukawas while fermions localised near the IR brane map to mainly composite states with correspondingly large Yukawas. 

In the usual picture, the bulk mass parameter is assumed to be constant. On the other hand, it is quite well motivated to consider that this bulk mass is dynamical and generated by coupling the fermions to a bulk scalar field which in turn obtains a VEV. 
We can then expect a position-dependent bulk mass as this VEV is generically not constant along the extra dimension.
In fact, the simplest mechanism for radion stabilisation, due to Goldberger and Wise \cite{Goldberger:1999uk}, consists in introducing a bulk scalar field which obtains a VEV from potentials on the two branes.
The most minimal scenario to dynamically generate the bulk mass is then to use this bulk scalar.
Interestingly, during the process of radion stabilisation, the profile of the Goldberger-Wise scalar VEV changes. When the latter is coupled to the fermions, this induces a change in the bulk masses of the fermions which in turn affects their wavefunction overlap with the Higgs on the IR brane and thus the Yukawa couplings. 
The RS model with bulk fermions therefore naturally allows for a scenario of varying Yukawa couplings during the EW phase transition.
Our goal is to study the cosmological dynamics of Yukawa couplings in this context.

The emergence of the EW scale in RS models comes during the stabilisation of the size of the AdS$_5$ slice. 
At high temperatures, the thermal plasma deforms the geometry and the IR brane is replaced by a black hole horizon. 
Going to lower temperatures, eventually a phase transition takes place and the IR brane re-emerges. This phase transition is typically strongly first-order and proceeds via bubble nucleation. 
The walls of these bubbles then interpolate between AdS$_5$ with an IR brane at infinity and at a finite distance. 
In the dual 4D theory, this transition is described by the dilaton -- which maps to the radion -- acquiring a VEV. 
To realise a model where the Yukawas are larger during the phase transition (as needed if we want to use the SM Yukawas as the unique $CP$-violating source during EW baryogenesis \cite{dynamicalyukawas,Berkooz:2004kx}),  we ask for the Yukawas to become larger when the IR brane is pushed to infinity.

We will discuss two realisations of this.
One way to induce varying Yukawas is to add an operator on the IR brane that effectively changes the value of the Yukawa coupling as the position of the IR brane changes.
This mechanism enables variations of order one for the Yukawas and can be relevant for $CP$-violation if applied to the top quark.
We discuss this option in sec.~\ref{sec:ModelI}.
The other possibility to implement large Yukawas during the phase transition  is to have a bulk mass for the fermions which decreases towards the IR. Since smaller bulk masses make the wavefunctions grow faster towards the IR, this leads to fermions which become increasingly IR-localized when the IR brane is pushed to infinity. The wavefunction overlap with the Higgs near the IR brane and thus the Yukawas then grow too. 
This mechanism can be relevant for $CP$-violation for all quarks and enables a large variation of the Yukawa couplings, from values of order one to today's small values of the light quarks. This realisation will be described in sec.~\ref{sec:ModelII}.

The plan of the paper is the following.
The motivations for this study are reported in sec.~\ref{sec:motiv}, where we summarise the key features of electroweak baryogenesis.
The Goldberger-Wise  mechanism  and  the description of the EW phase transition in RS models are reviewed  in secs.~\ref{sec:ReviewGW} and \ref{sec:EWPTRS} respectively. The derivation of 4D Yukawa couplings in RS models is reviewed in
sec.~\ref{sec:FlavourinRS}.
In sec.~\ref{sec:ModelI}, we present a first possible mechanism for Yukawa coupling variation, which relies on a new contribution to the Yukawa coupling on the IR brane. Sec.~\ref{sec:ModelII} discusses a generic mechanism for modifying fermionic profiles. The main idea is presented through a simple model in \ref{sec:GWprofile}. Its realistic implementation is given in sec.~\ref{sec:modifiedGW}.
In sec.~\ref{sec:flavourconstraints}, we discuss the implications of our constructions for flavour and $CP$-violating processes.
Sec.~\ref{sec:CFTint} provides the interpretation of the models in the dual CFT. We conclude in sec.~\ref{sec:conclusion}.

\section{Electroweak baryogenesis with varying Yukawa couplings}
\label{sec:motiv}

Electroweak baryogenesis is an appealing mechanism for explaining the baryon asymmetry of the universe, which relies on electroweak scale physics only (see e.g. \cite{Konstandin:2013caa}). It occurs in the framework of a first-order electroweak phase transition, in the vicinity of Higgs bubble walls, separating the broken phase where baryon number is conserved from the symmetric phase where sphaleron transitions are unsuppressed. 
Because of $CP$-violating interactions in the bubble walls between particles in the plasma and the Higgs, a chiral asymmetry may be generated and  converted into a baryon asymmetry by sphalerons in front of the bubble walls.
Due to the wall motion, the baryon asymmetry diffuses into the broken phase, where sphalerons are frozen, and the asymmetry is not washed out.  All models of EW baryogenesis postulate the existence of a new $CP$-violating source beyond the CKM phase, as needed to explain the baryon asymmetry. This is typically strongly constrained by measurements of electric dipole moments (EDMs), see e.g.  \cite{Huber:2006ri}.
However, as studied in depth in \cite{dynamicalyukawas}, if Yukawa couplings vary across the bubble walls, this provides a source of $CP$-violation which is active at early times only, and therefore not in tension with EDM experimental bounds. This source scales like 
\begin{equation}
S_{\slashed{CP}} \, \propto \,  \mbox{Im}[V^{\dagger} \, {M^{\dagger}}^{\prime \prime}  M \, V]_{ii} \, ,
\label{CPfromdynCKM}
\end{equation}
where $M$ is the fermion mass  matrix, $V$ is the matrix that diagonalizes ${M^{\dagger}}M$, the derivative is with respect to the coordinate perpendicular to the bubble wall and only the diagonal elements of the matrix in brackets are relevant.
Such a term vanishes for the Yukawas in the SM as they are constant across the bubble wall.
On the other hand, it is conceivable to use the CKM matrix as the $CP$-violating source for EW baryogenesis if the Yukawa couplings vary at the same time as the Higgs is acquiring its VEV $\langle H \rangle$. In fact, the correct amount of baryon asymmetry is naturally obtained if the Yukawa couplings varied from values of order 1 in the symmetric phase ($\langle H \rangle \rightarrow 0$) to their present value in the broken phase ($\langle H \rangle \rightarrow v_{\rm EW}$) \cite{dynamicalyukawas}.
This observation is the driving motivation for this  study and we are interested in providing a natural realisation of such Yukawa coupling variation during the EW phase transition.

There are two ways to get enough $CP$-violation from the source term (\ref{CPfromdynCKM}). 
It is possible to rely on the top Yukawa coupling only, provided that its phase changes during the EW phase transition. Indeed, writing $m_t = |m_t(z)| \, e^{-i\theta(z)}$, we have 
\be
\label{SourceTopQuark}
\mbox{Im}[ {m_t^*}^{\prime \prime}  m_t ] \, \rightarrow \, {[|m_t|^2\theta^{\prime}]}^{\prime} \, .
\ee
This can happen naturally in models where the top Yukawa coupling receives two contributions of order one,
\be
Y(z)=y_c + y_v(z) \, e^{i\theta_i} \, ,
\label{eq:twocontributions}
\ee
with $y_c$ being constant, while $y_v$ is varying across the bubble wall and $\theta_i$ is some arbitrary initial complex phase. This setup generates an effectively varying phase $\theta_{\rm eff}(z)=\operatorname{arg}(Y(z))$.
Using the profile of the Higgs VEV across the bubble wall, we can trade the coordinate $z$ for $\langle H \rangle$. This thus leads to a phase which effectively varies as the Higgs field is rolling towards the minimum of its potential:
\be
\theta_{\rm eff}(\langle H \rangle)=\operatorname{arg}(Y(\langle H \rangle)) \, .
\ee
As shown in \cite{Fromme:2006wx,Espinosa:2011eu, dynamicalyukawas}, if the top Yukawa coupling had such a varying phase during the electroweak phase transition, this can explain the baryon asymmetry of the universe.

The other possibility is to have Yukawa couplings whose phases do not vary but whose absolute values change. As follows from eq.~\eqref{SourceTopQuark} with $\theta=\text{const.}$, in this case the source for one flavour vanishes and we thus need at least two flavours. 
These two situations are studied in depth in \cite{dynamicalyukawas}. Although the full calculation is presented in \cite{dynamicalyukawas}, it was shown that the top-charm system gives the dominant contribution.
Our two models I and II discussed in sections \ref{sec:ModelI} and \ref{sec:ModelII} of this
 paper illustrate these two cases.
To introduce these new findings, we need first to review several basic features of the 
physics of Randall-Sundrum models.

\section{Review of the Goldberger-Wise mechanism}
\label{sec:ReviewGW}

We now  review a key aspect of RS models known as the Goldberger-Wise mechanism \cite{Goldberger:1999uk}.
The general construction we consider is based on a slice of AdS$_5$ with metric 
\be
ds^2 \, = \, e^{-2 k y} \eta_{\mu \nu} dx^\mu dx^\nu \, - \, dy^2 
\ee
and branes at $y=0$ (UV/Planck brane) and $y=y_\ir$ (IR/TeV brane). The theory could be defined on an orbifold or an interval. In either case, we restrict the coordinate $y$ to the interval $[0,y_\ir]$ here and below.\footnote{To calculate integrals over $\delta$-functions on the boundaries, we first move the $\delta$-functions $\epsilon$ away from the brane into the interval, perform the integral and then send $\epsilon$ to 0 (e.g.~$\underset{\epsilon \rightarrow 0}{\lim}\int_0^{y_\ir} \hspace{-.1cm} f(y) \delta(y-\epsilon) \, dy$). } We assume that the Higgs is localized on the IR brane, whereas the fermions and gauge bosons live in the bulk. We also introduce a real scalar field $\phi$ in the bulk with potentials on the branes. Its action reads
\begin{align}
\label{actionGW}
& S \, \supset \, \int d^5 x \,\sqrt{g} \, \Bigl(\frac{1}{2} \partial_A \phi \, \partial^A \phi  -  \frac{m_\phi^2}{2} \phi^2  -  \delta(y) \, V_\uv  -  \delta(y-y_\ir) \, V_\ir \Bigr) \, ,\\
& V_\uv \, = \,  \lambda_\uv (\phi^2 - v^2_\uv)^2\, , \quad \quad \quad V_\ir \, = \, \lambda_\ir (\phi^2 - v^2_\ir)^2 \, .
\label{BoundaryPotentials}
\end{align}
All dimensionful parameters (in particular the AdS curvature scale $k$) are expected to be of order one in Planck units.
The potentials cause the scalar to develop a VEV with a profile along the extra dimension given by (see e.g.~\cite{Goldberger:1999uk})
\be 
\label{vevprofile}
\langle \phi \rangle \, = \, A \, e^{(4+\epsilon) k y} \, + \, B \, e^{-\epsilon k y} \, ,
\ee
where 
\be
\epsilon \equiv \sqrt{4+m_\phi^2/k^2}-2 \, .
\ee
The constants $A$ and $B$ are determined by the boundary conditions which read
\begin{align} 
\label{boundaryconditionUV}
k \bigl( (4+\epsilon) A - \epsilon \, B \bigr)- \frac{1}{2} \frac{d V_\uv}{d \phi}\Bigr|_{0} \,&  = \, 0 \, ,\\
k \left( (4+\epsilon) \sigma_\ir^{-(4+\epsilon)} A - \epsilon \, \sigma_\ir^{\epsilon} B \right)+ \frac{1}{2} \frac{d V_\ir}{d \phi}\Bigr|_{y_\ir}\, & = \, 0 \, , 
\label{boundaryconditionIR}
\end{align}
where the warp factor at the IR brane,
\be
\sigma_\ir\equiv e^{-k y_\ir}\, ,
\ee
defines the radion field. The aim is to stabilize the radion such that
$\sigma_\ir \times k \sim \, \text{TeV}$, which represents the effective cutoff scale on the IR brane (and therefore for the Higgs mass parameter).
In the limit of large couplings $\lambda_\ir$ and $\lambda_\uv$, the last term in eqs.~\eqref{boundaryconditionUV} and \eqref{boundaryconditionIR} dominates and we get $\langle\phi\rangle(0)=v_\uv$ and $\langle\phi\rangle(y_\ir)=v_\ir$. This in turn gives
\begin{align} 
\label{leadingA}
A \, & = \, \frac{v_\ir - v_\uv \sigma_\ir^\epsilon}{\sigma_\ir^{-(4+\epsilon)}-\sigma_\ir^\epsilon} \, \simeq \, v_\ir  \, \sigma_\ir^{4+ \epsilon} \, - \, v_\uv  \, \sigma_\ir^{4+2 \epsilon} \, ,\\
B \, & = \, v_\uv -A \, \simeq \, v_\uv \, ,
\label{leadingB}
\end{align} 
where we have assumed that $v_\ir$ and $v_\uv$ are of comparable size. The leading corrections to this in $\lambda_{\uv},\lambda_{\ir}$ and to zeroth order in $\epsilon$ are given by \cite{Goldberger:1999uk}
\begin{align}
\label{deltaA}
\delta A \, & \simeq \, - \frac{k}{\lambda_\ir v_\ir^2} \, A \, ,\\
\delta B \, & \simeq \, \left(\frac{k}{\lambda_\uv v_\uv^2} + \frac{k}{\lambda_\ir v_\ir^2} \right) \, A \, .
\label{deltaB}
\end{align}
We see that $\delta B$ is suppressed relative to $B$ by powers of the warp factor $\sigma_\ir$ and is thus always negligible. Furthermore, $\delta A$  can be neglected relative to $A$ for $\lambda_\ir v_\ir^2 \gg k$ which we will assume in the following.

The contribution of the scalar VEV to the potential energy depends on the size of the extra dimension. Integrating over the extra dimension, the resulting potential for the radion $\sigma_\ir$ is given by \cite{Goldberger:1999uk}
\be 
\label{RadionPotential}
V(\sigma_\ir) \, = \,  (4 + \epsilon) \, k A^2 (\sigma_\ir^{-(4+2 \epsilon)}-1 )+ \epsilon \, k B^2 (1-\sigma_\ir^{4+2\epsilon}) + V_\uv\bigl(\phi(0)\bigr)+ \sigma_\ir^4 V_\ir\bigl(\phi(y_\ir)\bigr) \, .
\ee
In the limit of large $\lambda_{\uv},\lambda_{\ir}$, the boundary conditions give $\langle\phi\rangle(0)=v_\uv$ and $ \langle\phi\rangle(y_\ir)=v_\ir$ and the boundary potentials thus vanish. The corrections to this coming from eqs.~\eqref{deltaA} and \eqref{deltaB} for finite $\lambda_{\uv},\lambda_{\ir}$ are negligible for $\lambda_\ir v_\ir^2 \gg k$. Similarly, the corrections to the $A^2$- and $B^2$-dependent terms in eq.~\eqref{RadionPotential} are then negligible too.
The potential has a minimum for $\epsilon >0$. Expanding in $\epsilon$, we find
\be
\label{hierarchyrelation}
\sigma_{\ir}^{\rm min} \, = \, \left(\frac{v_\ir}{v_\uv}\right)^{1/\epsilon} \left(  1 \, + \, \sqrt{\frac{\epsilon}{4}} \, + \, \mathcal{O}(\epsilon) \right)^{1/\epsilon} \, .
\ee
A large hierarchy can thus be generated from an $\mathcal{O}(1)$-ratio $v_\ir/v_\uv$ if $\epsilon\ll 1$. Note that an additional term $\delta T_\ir \sigma_\ir^4$ in the potential can allow for a minimum also for negative $\epsilon$ \cite{Creminelli:2001th,Randall:2006py}. Such a term can arise from a detuned brane tension on the IR brane. However, we find that in the two models that we consider negative $\epsilon$ causes the Yukawa couplings to become nonperturbative when the IR brane is sent to infinity.\footnote{For model I, this can be anticipated from eq.~\eqref{5dYukawa}. The new contribution to the Yukawa coupling remains proportional to $\sigma_\ir^\epsilon$ for negative $\epsilon$ which causes it to diverge in the limit $\sigma_\ir \rightarrow 0$. Model II with negative $\epsilon$ can give growing wavefunctions for decreasing $\sigma_\ir$ if $\tilde{c} < 0 < c$ for the modified profile discussed in sec.~\ref{sec:modifiedGW}. For sufficiently small $\sigma_\ir$, this results in the fermions being IR-localized. Using eq.~\eqref{EnApprox}, we see from eq.~\eqref{ModifiedYukawaCouplingsMod} that the Yukawa coupling then is proportional to factors of $\sqrt{1-2c -2 \tilde{c} \sigma_\ir^\epsilon}$ for each of the two fermions. Again this diverges in the limit $\sigma_\ir \rightarrow 0$.} We therefore focus on positive $\epsilon$ in this paper. 

Note that the bulk potential in eq.~\eqref{actionGW} only contains a mass term for $\phi$. In principle also higher-order terms in $\phi$ can appear. The leading such term, $\phi^3$, was included in the analysis of refs.~\cite{Chacko:2013dra,Chacko:2014pqa}. The resulting profile for the Goldberger-Wise scalar was found to have $\mathcal{O}(1)$-corrections compared to the profile for a bulk potential with only a mass term. Note that such a $\phi^3$-term is in principle expected in model II discussed later because of the Yukawa coupling in the bulk, though it may be small.
Nevertheless even if it is sizeable the profile for positive $\epsilon$ will still decay by an $\mathcal{O}(1)$-factor when going from the UV to the IR. This is the crucial feature that we need for model II and we therefore expect this mechanism to work also if the bulk potential contains higher-order terms in $\phi$. It is less clear, on the other hand, if the derivative of the Goldberger-Wise scalar at the IR brane is then still suppressed when the radion is at the minimum of its potential. This is the crucial feature which is  needed for model I discussed later. However, as it has no bulk Yukawa coupling, the $\phi^3$-term in model I can be forbidden by imposing a $\mathcal{Z}_2$-symmetry.   
We leave a detailed study of our mechanism for this more general case to future work.

The non-constant piece of  the potential (\ref{RadionPotential}) is of the dilaton type,
\be
V(\sigma_\ir) \, \sim \sigma_\ir^4 \times f(\sigma_\ir^{\epsilon}) \, ,
\ee
where $f$ is a very slowly-varying function since it depends on  $\sigma_\ir^{\epsilon}$ only.
The cosmological dynamics of this very shallow  potential was summarized in ref.~\cite{Konstandin:2011dr}.
We discuss it next.

\section{The electroweak phase transition in Randall-Sundrum models}
\label{sec:EWPTRS}

While there has been an extensive literature on the phenomenology of Randall-Sundrum models, little has been established on its early cosmology.
On the other hand, the attractivity of this solution to the hierarchy problem also relies on whether it is cosmologically realistic.
One of the very first aspects to be checked was that the Friedmann equation could in fact be recovered, as expected, since gravity is effectively 4-dimensional in this model, at energies below the EW scale when the radion is stabilized \cite{Cline:1999ts,Csaki:1999jh}.

\begin{figure}[t]
\centering
\includegraphics[width=10cm]{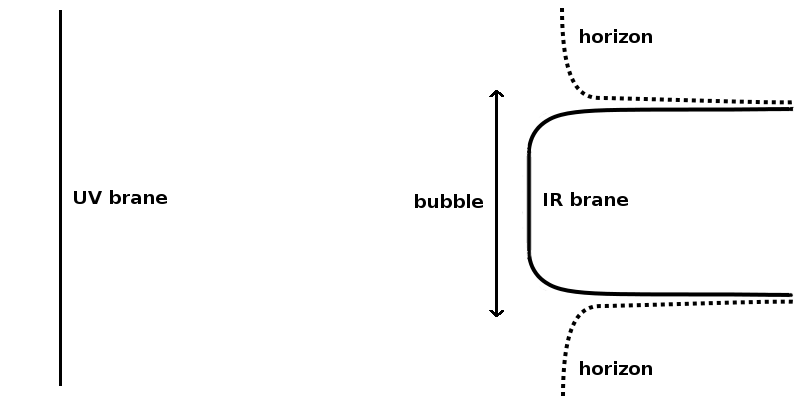}
\caption{\label{fig:PhaseTransition} \small
Schematic depiction of the phase transition: A bubble of the Randall-Sundrum phase emerges from the surrounding AdS-Schwarzschild phase. In the transition region between the two phases, one sees the black hole horizon receding to infinity and subsequently the IR brane coming in from infinity. }
\end{figure}

On the other hand, the knowledge of what happened before radion stabilisation is less under control. Nevertheless the phase transition leading to the stabilisation of the radion can be understood as follows \cite{Creminelli:2001th}:
At high temperatures, the system is in an AdS-Schwarzschild phase with a UV brane and a black hole horizon in the IR (whose Hawking temperature matches the temperature of the system). In the dual picture, this corresponds to the strongly-coupled theory being in the deconfined phase and the free energy scales like $F_{\mbox{\tiny  AdS--S}} \propto - T^4$ as expected. Going to lower temperatures, eventually a phase transition happens and the black hole horizon is replaced by the IR brane. This phase transition is typically strongly first-order and proceeds via bubble nucleation. Both geometries -- AdS-Schwarzschild and the Randall-Sundrum geometry with two branes -- have different topologies. They can be smoothly connected, however, by sending respectively the horizon and the IR brane to infinity which gives pure AdS$_5$ (cutoff by the UV brane). It is therefore expected that the bubble walls interpolate between the two phases as follows \cite{Creminelli:2001th}: Going perpendicular to the bubble wall from the AdS-Schwarzschild phase outside towards the Randall-Sundrum phase inside, we first see the horizon receding until we arrive at pure AdS$_5$. Further towards the inside, the IR brane comes in from infinity until it arrives at its stabilized position as determined by the Goldberger-Wise mechanism. This is depicted in fig.~\ref{fig:PhaseTransition}. The radion $\sigma_\ir$ thus varies from $0$ on the outside of the bubble wall to $\sigma_\ir^{\rm min}$ inside the bubble. This behaviour will be crucial for us as our models have Yukawa couplings which grow with decreasing $\sigma_\ir$ and thus grow across the bubble walls.

This phase transition to the Randall-Sundrum phase is special due to the nearly conformal nature of the radion potential (\ref{RadionPotential}) whose  cosmology is different from the one of usual polynomial scalar potentials. 
The tunneling action can be calculated by determining the bounce solution for the radion potential which was given in the previous section
\cite{Creminelli:2001th,Randall:2006py,Nardini:2007me,Konstandin:2010cd,Konstandin:2011dr}.
One finds that in the calculable region of parameter space, the phase transition may complete but is typically very strong and happens after significant amount of supercooling, see \cite{Konstandin:2011dr} and \cite{Caprini:2015zlo} for a recent updated status summary.\footnote{See however ref.~\cite{Hassanain:2007js} for an alternative solution changing this conclusion.}
Indeed the nucleation temperature can be parametrically much smaller than the scale associated with the minimum of the potential. 
An interesting signature of this scenario is the typical large signal amplitude of the stochastic gravity wave background peaked in the millihertz range inherited from the time of the phase transition, and observable at LISA \cite{Randall:2006py,Caprini:2015zlo}.

\begin{figure}[t]
\centering
\includegraphics[width=16cm]{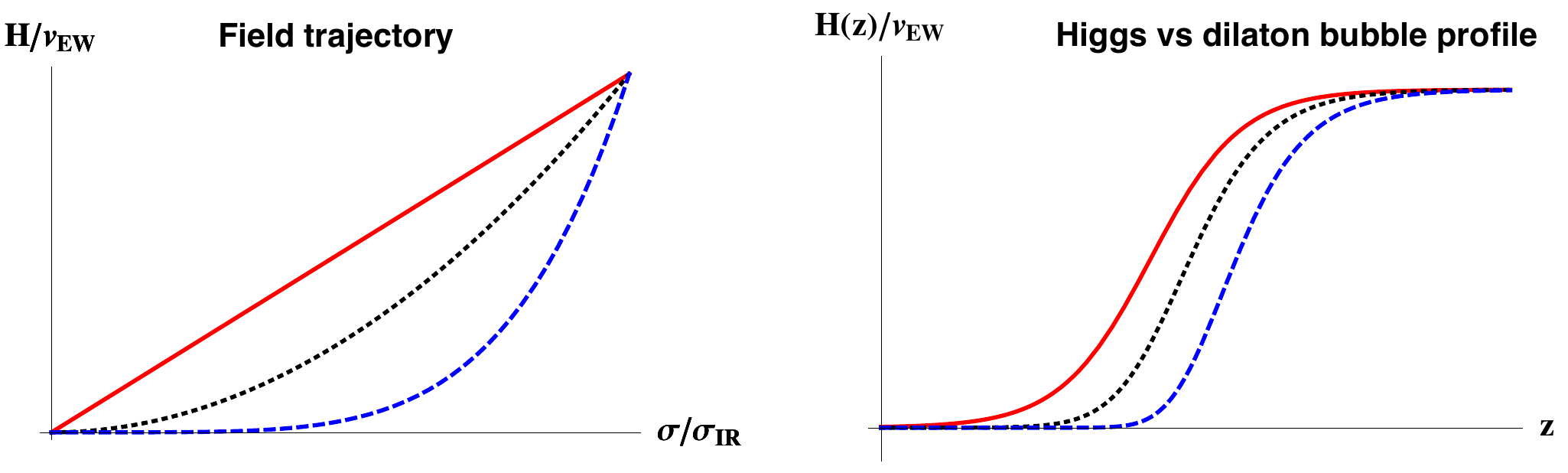}
\caption{\label{fig:higgsversusdilaton} \small
Left: Sketch of possible paths in the dilaton-Higgs
field space. The red solid line reflects the linear relationship (\ref{RelationHiggsVEVRadionVEV}) between the two VEVs. Right: The corresponding profiles of the fields along the bubble walls. The red solid line corresponds to the dilaton and Higgs bubble walls precisely overlapping and thus the linear relationship (\ref{RelationHiggsVEVRadionVEV}). For the dotted and dashed lines,  the dilaton reaches its minimum before the Higgs does, which tends to attenuate the variation of the Yukawa couplings during the EW phase transition. }
\end{figure}

During the phase transition, also the electroweak symmetry gets broken. We assume that the Higgs is localized on the IR brane. The action for the Higgs then reads
 \be
 \label{HiggsKineticTermPotential}
S \, \supset \, \int d^5 x \,\sqrt{g} \,  \delta(y-y_\ir)  \left( e^{2ky} \, \eta^{\mu \nu} D_{\mu} \tilde{H}^{\dagger} D_{\nu} \tilde{H} 
 - \lambda  \left( |\tilde{H}|^2-v_P^2 \right)^2 \right),
 \ee
 where $v_P\sim M_{\rm Pl}$. In terms of the canonically normalised Higgs field $H=\sigma_\ir \tilde{H}$, the Higgs potential reads
 \be
 V(H)= \lambda  \left( |{H}|^2-v_P^2 \,  \sigma_{\rm IR}^2 \right)^2 \, .
 \ee
 The Higgs VEV then scales like
 \be
 \label{RelationHiggsVEVRadionVEV}
\langle H \rangle = v _{\rm EW} \times \frac{\sigma_\ir}{\sigma_\ir^{\rm min}} \, ,
 \ee
where $v _{\rm EW}=v_P \sigma_\ir^{\rm min}=174 \,\text{GeV}$ is the electroweak scale. 
In deriving eq.~\eqref{RelationHiggsVEVRadionVEV}, we have assumed that the Higgs is always at the minimum of its potential during the phase transition to the Randall-Sundrum phase.
This is an idealised situation, however, we can expect this description to be physically sensible.
To derive the exact relation between the Higgs VEV and the radion VEV, one  has to compute the bounce, something which we postpone to future work.
The special features in the RS case are i) a nearly conformal potential along the radion direction, ii)  no quadratic term for the Higgs if the radion vanishes.
Therefore, electroweak symmetry cannot be broken unless the radion has a VEV. 
If electroweak symmetry breaking takes place only  after the radion settled in its minimum, then there is no variation of the Yukawa couplings during the electroweak phase transition. It is therefore crucial that they both change at the same time, which is what we expect. Indeed, if the Higgs and the radion were on equal footing, i.e.~both having similar potentials, then the path in the two-dimensional field space would be along the diagonal as both fields would follow the same tendency if they have similar masses. On the other hand, if the radion is much heavier than the Higgs, we expect the tunneling to proceed first along the radion direction and only then along the Higgs direction. 
We illustrate this schematically in fig.~\ref{fig:higgsversusdilaton}.
Therefore, the optimal case will be for a relatively light radion. Determining the precise relation between the Higgs and radion VEVs as a function of the radion mass will be an interesting task in itself.

The breaking of electroweak symmetry is thus tied to the radion cosmology.\footnote{Implications for cold baryogenesis were studied in refs.~\cite{Konstandin:2011ds,Servant:2014bla}.} Since the phase transition to the Randall-Sundrum phase is typically strongly first-order, the electroweak phase transition is then first-order too. This motivates the possibility of electroweak baryogenesis, provided that the bubble wall velocity is smaller than the sound speed (for larger velocities the baryon asymmetry vanishes as there is no time for $CP$-violating diffusion processes in front of the bubble walls where sphalerons are active).
In fact, the danger for electroweak baryogenesis  in  strong first-order phase transitions is that the friction exerted by the plasma on the wall might not be sufficient to prevent the bubble wall from a runaway behavior in which case the wall keeps accelerating, towards ultra-relativistic velocities.
The determination of the bubble wall velocity is a non-trivial calculation. It depends on the strength of the phase transition, i.e.~the amount of latent heat released, as well as the amount of friction between the particles in the plasma and the bubble wall  \cite{Espinosa:2010hh}. Friction is due to particles changing mass across the wall.
In contrast with the SM or the MSSM, we expect that a large number of degrees of freedom become very massive during  
the RS phase transition. Since  the precise theory in the CFT phase is unknown (in particular the number of CFT degrees of freedom), the friction  is left as a free parameter. But we can expect that for a large number of CFT degrees of freedom, friction will be relevant. It is clear however that it will be effective only for not too low nucleation temperatures.
As the nucleation temperature is typically smaller than the scale set by the radion VEV at the minimum of its potential \cite{Konstandin:2011dr}, conditions for EW baryogenesis may not be satisfied for a generic  choice of parameters.
We leave the model-dependent detailed analysis of the EW phase transition for future work.
 Therefore it should be clear that the possibility of EW baryogenesis is based on the assumption that there exists a region in parameter space where the phase transition is moderately strong and the bubble wall velocity can be subsonic. We then show that the RS setup generically incorporates the variation of Yukawas during the EW phase transition and therefore enables to realise EW baryogenesis with the CKM matrix as the only $CP$-violating source.\footnote{Note that another paper, ref.~\cite{Perez:2005yx}, entertained the idea of varying Yukawas during the dynamics  that stabilize fermion profiles in (unwarped) extra-dimensional models, however, at a scale much above the electroweak scale, and therefore for a baryogenesis mechanism requiring 
higher-dimensional $B-L$- violating operators.}

Note that even if the bubble wall velocity is supersonic, our discussion is relevant since baryogenesis at the electroweak scale is still possible through a different mechanism, so-called ``cold baryogenesis", which does not rely on a transport mechanism, and is especially motivated  in the context of the supercooled RS phase transition  \cite{Konstandin:2011ds,Servant:2014bla}.
The source of $CP$-violation that we find from Yukawa variation could be used also in this context.

We thus want to generate Yukawa couplings between the Higgs and the fermions that change in size when the IR brane is moved away from the minimum of the Goldberger-Wise potential. 
To this end, we consider in  sections \ref{sec:ModelI} and \ref{sec:ModelII} two realisations, first through a new IR contribution from the Goldberger-Wise field to the Yukawa couplings and second through the bulk coupling of the Goldberger-Wise field to the fermions. Before doing that, we review how Yukawa couplings arise in RS models.

\section{Review of fermions in Randall-Sundrum models}
\label{sec:FlavourinRS}
We now review fermions in RS models and how the fermion mass hierarchy arises. 
In this paper, we are mainly interested in the Yukawa couplings of the up-type quarks. We denote by $\mathcal{Q}$ and $\mathcal{U}$ the bulk fields that give rise to the left-handed quark doublet and the right-handed up-type quarks, respectively. Including the kinetic term for completeness, the bulk action for the left-handed quark doublets $\mathcal{Q}$ reads (see e.g.~\cite{Grossman:1999ra})
\be 
\label{fermionaction}
S \, \supset \, \int d^5 x \sqrt{g} \, \left( E^A_a \left[\frac{i}{2} \, \overline{\mathcal{Q}} \, \gamma^a \left(\partial_A - \overleftarrow{\partial_A} \right) \mathcal{Q} \, + \, \frac{\omega_{bcA}}{8} \,  \overline{\mathcal{Q}} \{ \gamma^a ,\sigma^{bc} \} \mathcal{Q} \right] \, + \, c_\mathcal{Q} \, k \, \overline{\mathcal{Q}} \mathcal{Q} \right)  
\ee
and similarly for the right-handed up-type quarks $\mathcal{U}$.\footnote{On 
an orbifold $c_\mathcal{Q}$ needs to be odd, $c_\mathcal{Q} \propto \operatorname{sgn}(y)$, since $\bar{\mathcal{Q}} \mathcal{Q}$ is odd. Alternatively we can define the theory on an interval and then impose the same boundary conditions as on the orbifold.} 
$E^A_a$ is the inverse vielbein, $\omega_{bcA}$ is the spin connection and $c_\mathcal{Q} k$ with $c_\mathcal{Q}\sim {\cal O}(1)$ is the bulk mass of the 5D fermion. For simplicity, we have suppressed the flavour indices. Note that we can perform unitary transformations such that the kinetic terms and the mass terms are diagonal in flavour space. We will use this basis throughout this paper. The Yukawa coupling reads
\be
\label{localizedYukawa}
S \, \supset \, \int d^5x \, \sqrt{g} \; \delta(y-y_\ir) \lambda_u \, \tilde{H} \bar{\mathcal{Q}} \, \mathcal{U}  \, + \, \text{h.c.} \, ,
\ee
where $\lambda_u$ has dimension -1 and $\tilde{H}$ is the brane-localized Higgs (whose kinetic term and potential are given in eq.~\eqref{HiggsKineticTermPotential}).

We decompose the bulk fermions $\mathcal{Q}$ and $\mathcal{U}$ into left- and right-handed spinors and Kaluza-Klein (KK) modes.
This gives $\mathcal{Q}_{L,R} \equiv \frac{1}{2}(1\mp \gamma_5)\mathcal{Q}$ and 
\be 
\mathcal{Q}_{L,R}(x,y) \, = \, \sqrt{k} \, \sum_n e^{2 k y} f_{L,R}^{(n)}(y) \, \mathcal{Q}_{L,R}^{(n)}(x) 
\ee
and similarly for $\mathcal{U}$. The equations of motion for $\mathcal{Q}$ then read
\be 
\label{fermioneom}
\bigl(\pm \partial_y + c_\mathcal{Q}(y)\, k \bigr) f_{L,R}^{(n)} \, + \, m_\mathcal{Q}^{(n)} e^{k y} f_{R,L}^{(n)} \, = \, 0 \, ,
\ee
where $m_\mathcal{Q}^{(n)}$ are the KK masses. Notice that we have allowed for the possibility that $c_\mathcal{Q}$ is a function of $y$ which will become important later.
The wavefunctions fulfill the orthonormality conditions\footnote{For the general case of a position-dependent bulk-mass parameter $c(y)$, the equations of motion for the two chiralities can be combined and rewritten as
$$- \partial_y \, p_{_{L,R}} \, \partial_y \, \tilde{f}^{(n)}_{L,R} = (m^{(n)})^2 \, e^{2 k y}  \, p_{_{L,R}} \, \tilde{f}^{(n)}_{L,R} \, ,$$
where 
$$\tilde{f}^{(n)}_{L,R} \equiv e^{\pm k \int_0^y d\tilde{y} \, c(\tilde{y})} f_{L,R}^{(n)}\quad \text{and} 
\quad p_{_{L,R}} \equiv e^{-k y} e^{\mp 2 k \int_0^y d \tilde{y} \,  c(\tilde{y})} \, . $$
This has the form of a Sturm-Liouville equation (see e.g. eq.~(13) in ref.~\cite{Gherghetta:2010cj}). The problem therefore has a discrete set of real eigenvalues $(m^{(n)})^2$. The eigenfunctions $\tilde{f}_{L,R}^{(n)}$ form a complete set and satisfy the orthonormality relation
$$ \int_0^{y_\ir} dy \,k \, e^{2 k y}  \, p_{_{L,R}} \, \tilde{f}_{L,R}^{(n)} \, \tilde{f}_{L,R}^{(m)} \, = \, \delta_{nm} \, $$
which gives eq.~\eqref{orthonormalityconditions}. This guarantees that the Lagrangian in terms of the KK modes is diagonal. 
}
\be 
\label{orthonormalityconditions}
\int^{y_\ir}_0 \hspace*{-.2cm} dy \, e^{k y} k \, f_{L,R}^{(m)} \, f_{L,R}^{(n)}  \, = \, \delta_{m n} \, .
\ee
In order to ensure that the boundary terms vanish after the variation of the action, we can impose that either the left- or right-handed fermion is zero at the two branes (see e.g.~\cite{Csaki:2003sh}). This leaves one chiral massless mode, $m_\mathcal{Q}^{(0)}=0$, which we identify with the SM fermion. We then choose the boundary conditions such that $\mathcal{Q}$ has a left-handed massless mode, whereas the massless mode from $\mathcal{U}$ is right-handed.  
If the bulk masses $c_\mathcal{Q} k$ and $c_\mathcal{U}k$ are constant, as usually assumed in the literature, 
the wavefunctions for the left-handed massless modes from $\mathcal{Q}$ then read
\be 
\label{righthandedmasslessmode}
f_L^{(0)}(y)\, = \, \mathcal{N}_{{c}_\mathcal{Q}}^{(0)} \, e^{-c_\mathcal{Q} k y} \, ,
\ee
where 
\be 
\label{normalizationmasslessmode}
\mathcal{N}^{(0)}_{{c_\mathcal{Q}}} \, = \, \sqrt{\frac{1-2 c_\mathcal{Q}}{\sigma_\ir^{2 c_\mathcal{Q} -1}-1}}
\ee
is a normalization constant. For later convenience, we redefine ${c} \rightarrow -{c}$ for the bulk fermions $\mathcal{U}$ with right-handed massless modes.
Their wavefunctions $f_R^{(0)}(y)$ are then again given by eqs.~\eqref{righthandedmasslessmode} and \eqref{normalizationmasslessmode} with ${c}_\mathcal{Q}$ replaced by ${c}_\mathcal{U}$. With this convention, both left- and right-handed massless modes are UV (IR) localized for $c>1/2$ ($c<1/2$).

The effective 4D Yukawa coupling between the SM fermions and the Higgs is given by 
\be 
\label{Action4dYukawa}
S \, \supset \,\int d^4 x \,  y_{u}(\sigma_\ir) \, H \, \bar{\mathcal{Q}}^{(0)}_{L} \, \mathcal{U}^{(0)}_{R} \, + \, \text{h.c.} \, ,
\ee
where $H \equiv \sigma_\ir \tilde{H}$ to obtain a canonically normalized kinetic term and
\be
\label{4dYukawa}
y_{u}(\sigma_\ir) \, \equiv \,{\lambda}_u \,k \, \sqrt{\frac{1-2c_\mathcal{Q}}{1-\sigma_\ir^{1-2c_\mathcal{Q}}}}\,\sqrt{\frac{1-2c_\mathcal{U}}{1-\sigma_\ir^{1-2c_\mathcal{U}}}} \, .
\ee
For $c_\mathcal{Q},c_\mathcal{U}> 1/2$, this becomes exponentially suppressed. This shows how large hierarchies between the 4D Yukawa couplings can be obtained in RS starting from bulk mass parameters and 5D Yukawa couplings of order one in units of the AdS scale $k$. Notice that already in this setup the Yukawa couplings depend on the position $\sigma_\ir$ of the IR brane. Since the light quarks are all localized towards the UV brane, however, their Yukawa couplings decrease when the IR brane is sent to infinity, $\sigma_\ir \rightarrow 0$. Correspondingly, they are small in a large portion of the bubble wall during the phase transition and $CP$-violation is suppressed. We will later see how modified fermion profiles can lead to increased Yukawa couplings during the phase transition. 

The parameters that determine $y_u$ need to be chosen such that the measured masses and mixing parameters are reproduced. This still leaves a considerable freedom. For definiteness, we will use a benchmark point for these parameters from ref.~\cite{Casagrande:2008hr}. 
We need to adjust the parameters, however, since for the benchmark point a hierarchy $\sigma_\ir^{\rm min}=10^{-16}$ was assumed, whereas we choose $\sigma_\ir^{\rm min}=2.5 \cdot 10^{-15}$ in this paper.\footnote{For example for $k \sim M_5 \sim M_{\rm Pl}$, this would give an IR scale $k \sigma_\ir^{\rm min} \sim 5 \, \text{TeV}$. This would be consistent with electroweak precision tests even without a custodial symmetry (though it requires a cancellation of order 25\% in the contributions to $\epsilon_K$ to be viable) \cite{Malm:2013jia,Bauer:2016lbe}.} In addition, we reduce the 5D Yukawa couplings involving the left-handed top-bottom doublet by a factor $3/8$ compared to those of the benchmark point. This will ensure that the couplings do not become nonperturbative in the limit $\sigma_\ir \rightarrow 0$ in the models that we consider later. After making these two modifications, we adjust the bulk-mass parameters such that the 4D Yukawa couplings are again reproduced. We will only list the parameters for the top-charm sector since it gives the dominant effect for the models that we consider later. In a basis such that the couplings in the Lagrangian are proportional to $(\bar{\mathcal{Q}}_2,\bar{\mathcal{Q}}_3) \lambda_u (\mathcal{U}_2,\mathcal{U}_3)^T$, the 5D Yukawa couplings read
\be 
\label{BenchmarkYukawaCouplings}
{\lambda}_u \, = \, \frac{1}{k} \, \begin{pmatrix} 0.76 \cdot e^{-1.46 i} & 0.74 \cdot e^{-2.13 i} \\ 0.28 \cdot e^{0.39 i} & 0.93 \cdot e^{-1.26 i} \\ \end{pmatrix} 
\ee
and the bulk-mass parameters are
\be
\label{BenchmarkPointParameters}
c_{\mathcal{Q}_2} \, = \, 0.521 \qquad \quad c_{\mathcal{U}_2} \, = \, 0.565 \qquad \quad c_{\mathcal{Q}_3} \, = \, 0.278 \qquad \quad c_{\mathcal{U}_3} \, = \, -0.339 \, .
\ee
Here and below indices on the fields $\mathcal{Q}$ and $\mathcal{U}$ denote the generation. The value for $c_{\mathcal{Q}_3}$ can be consistent with constraints from the $Zb \overline{b}$-coupling \cite{Casagrande:2008hr}. Together with the other parameters for the benchmark point, these parameter values reproduce the measured quark masses and mixings when the running from an IR scale of $1.5 \, \text{TeV}$ to the electroweak scale is taken into account. Note that we assume a slightly larger IR scale. However, we expect the required adjustments in the parameters that we are interested in to be small and will neglect them in the following. Note also that the Yukawa couplings in the plots are thus given at $1.5 \,\text{TeV}$ and will change slightly when run down to the electroweak scale. 

In the next sections, we will make some rather small but influential modifications to this commonly used picture.

\section{Model I: A new IR contribution to the Yukawa couplings}
\label{sec:ModelI}

\begin{figure}[t]
\centering
\includegraphics[width=8cm]{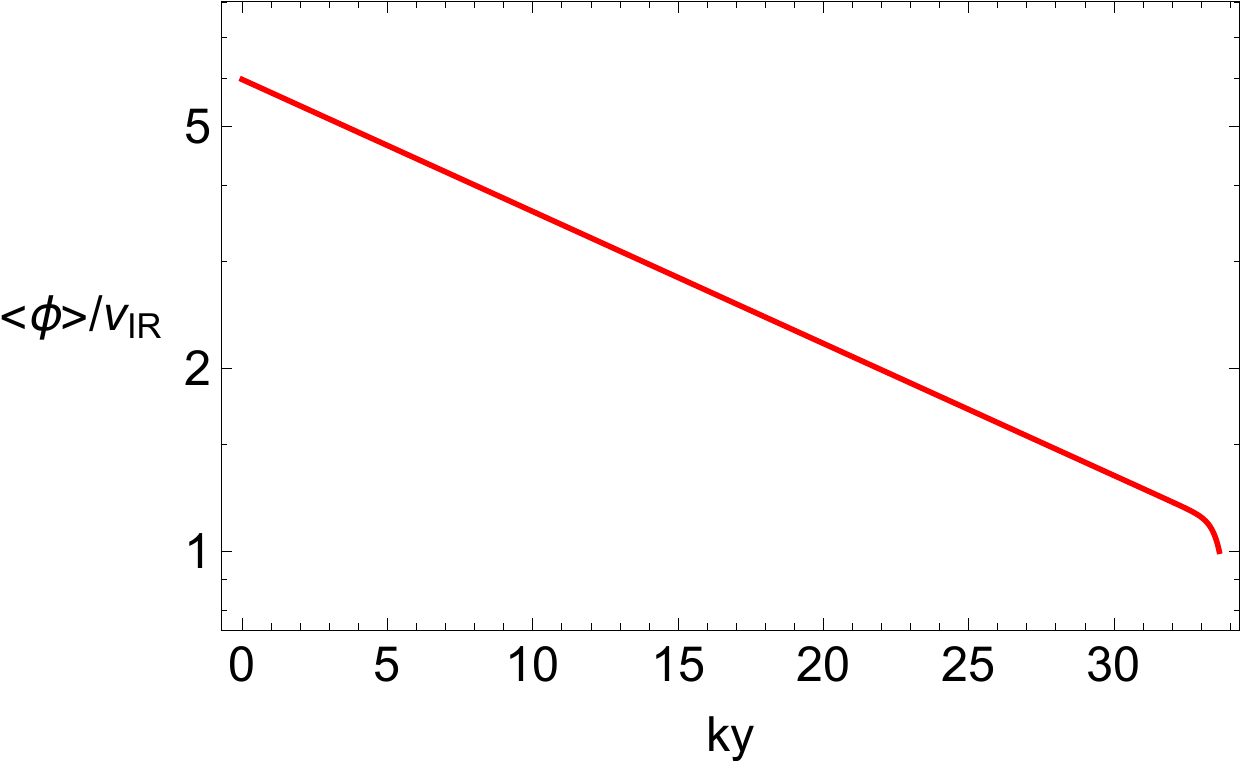}
\includegraphics[width=8cm]{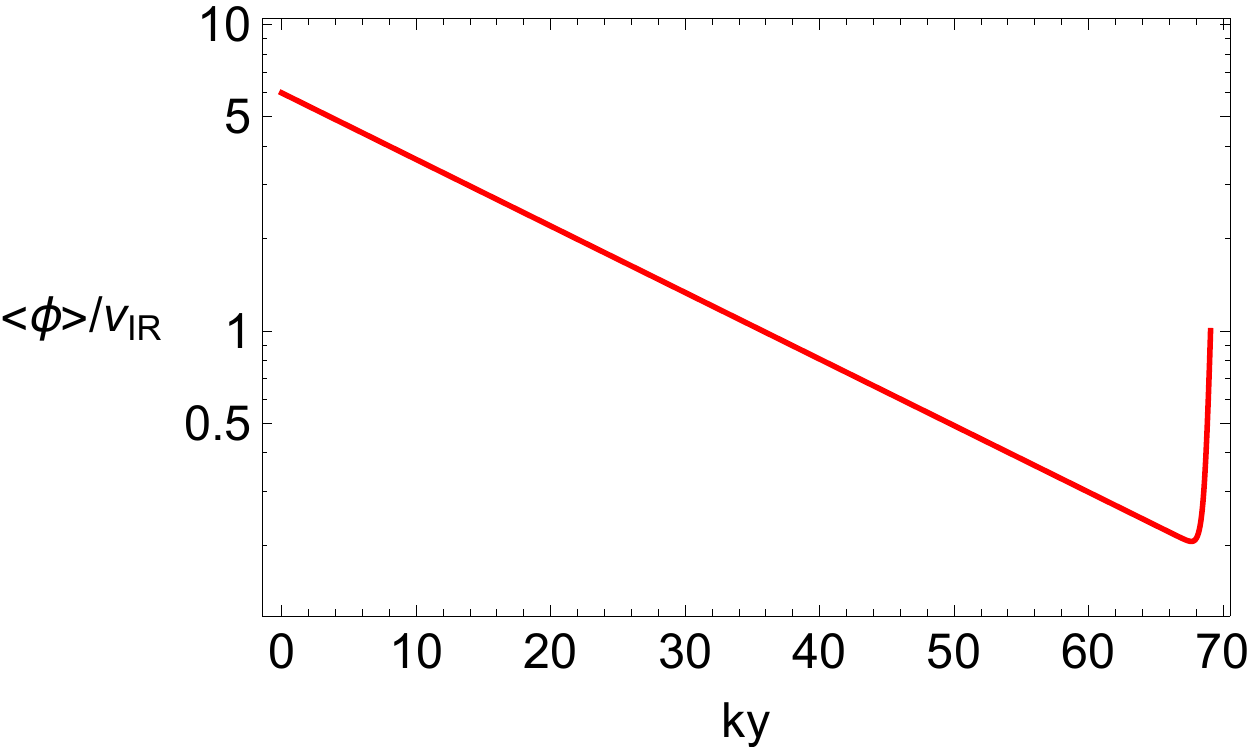}
\caption{\label{fig:GWprofile} \small
(a) VEV of the Goldberger-Wise scalar in eq.~\eqref{vevprofile} along the extra dimension if the radion is at the minimum of its potential, chosen as $\sigma_\ir^{\rm min}=e^{-ky^{\rm min}_\ir} = 2.5\cdot 10^{-15}$. (b) VEV for the same parameters as in (a) but for the radion at $\sigma_\ir  =10^{-30}$.}
\end{figure}

The first model that we present involves a higher-dimensional coupling of the Goldberger-Wise scalar to the Yukawa operator $\tilde{H} \bar{\mathcal{Q}} \, \mathcal{U}$ on the IR brane. This gives an additional contribution to the Yukawa coupling. We then use the fact that the VEV of the Goldberger-Wise scalar changes when the IR brane is moved, leading to a change in the Yukawa coupling. The boundary potential keeps the VEV at the IR brane relatively constant, $\langle \phi \rangle(y_\ir)\simeq v_\ir$. A coupling $\phi \tilde{H} \bar{\mathcal{Q}} \, \mathcal{U}$ therefore does not result in a sufficient change for our purposes. We instead consider a derivative coupling which can for example arise due to a finite thickness of the brane.
For the up-type quarks, eq.~(\ref{localizedYukawa}) now becomes
\be
\label{NewYukawaCoupling}
S \, \supset \, \int d^5x \, \sqrt{g} \; \delta(y-y_\ir) \left(\lambda_u \, \tilde{H} \bar{\mathcal{Q}} \, \mathcal{U}\, + \, \kappa_u \, \partial_y \phi \, \tilde{H} \bar{\mathcal{Q}} \, \mathcal{U} \right) \, + \, \text{h.c.} \, .
\ee
We have again suppressed the flavour indices for the fields and the coupling constants $\lambda_u$ and $\kappa_u$ (which have dimensions $-1$ and $-7/2$, respectively). Similar couplings can exist for the down-type quarks but it is enough to focus on the up-type couplings for our purposes. 
The nonvanishing derivative of the VEV \eqref{vevprofile} of the Goldberger-Wise scalar at the IR brane gives an additional contribution to the 5D Yukawa coupling which depends on the position of the IR brane: 
\be
S \, \supset \, \int d^5x \, \sqrt{g} \; \delta(y-y_\ir) \,\tilde{\lambda}_u(\sigma_\ir) \, \tilde{H} \bar{\mathcal{Q}} \, \mathcal{U}\, + \, \text{h.c.} \, ,
\ee
where
\be 
\label{5dYukawa}
\tilde{\lambda}_u(\sigma_\ir) \, \simeq \, \left[\lambda_u  \, + \, 4 \, \kappa_u  k\,  v_\ir \left[1-\left(1+\sqrt{\frac{\epsilon}{4}} \right)\left(\frac{\sigma_\ir}{\sigma_\ir^{\rm min}}\right)^\epsilon \right] \right] \, .
\ee
Note that the contribution from the derivative coupling is suppressed by a factor $\sqrt{\epsilon}$ if the radion is at the minimum of its potential, $\sigma_\ir =\sigma_\ir^{\rm min}$. This can be understood as follows: Both the bulk potential $m_\phi^2 \phi^2$ and the kinetic term $(\partial_y \phi)^2$ of the Goldberger-Wise scalar in eq.~(\ref{actionGW}) contribute to the radion potential. Since $ m_\phi^2 \simeq 4k^2 \epsilon$, the former is suppressed by $\epsilon$. The minimum of the potential then occurs at a radion VEV for which the latter is suppressed by $\epsilon$ too. This leads to
\be
\partial_y \langle\phi\rangle \propto \sqrt{\epsilon}
\ee
near the stable position of the IR brane. This suppression can be seen in fig.~\ref{fig:GWprofile}(a), where we plot the VEV of the Goldberger-Wise scalar along the extra dimension if the radion is at the minimum of its potential (we choose $\sigma_\ir^{\rm min}=2.5 \cdot 10^{-15}$ and $\epsilon =1/20$).
The suppression is lifted when the IR brane is moved to infinity, $\sigma_\ir \rightarrow 0$, and the Yukawa coupling correspondingly grows. This is visible in fig.~\ref{fig:GWprofile}(b) which shows the VEV for the same parameters as in fig.~\ref{fig:GWprofile}(a) but with the radion at $\sigma_\ir=10^{-30}$.

The resulting 4D Yukawa coupling is obtained from eq.~(\ref{4dYukawa}) with the replacement $\lambda_u \rightarrow \tilde{\lambda}_u$.
The new contribution to the effective 5D Yukawa coupling $\tilde{\lambda}_u$ grows by a factor $\sqrt{4/\epsilon}$ when $\sigma_\ir$ is changed from $\sigma_\ir^{\rm min}$ to $0$. Accordingly, this model enables variations in the Yukawa couplings of order one only. The Yukawa coupling receives two contributions like in eq.~(\ref{eq:twocontributions}), on the other hand, and we can therefore still use it for the top quark as discussed in sec.~\ref{sec:motiv}. 
Note that since the top is localized in the IR, the prefactor from the wavefunction overlaps in eq.~\eqref{4dYukawa} depends only very weakly on $\sigma_\ir$ (for the bulk mass parameters in eq.~\eqref{BenchmarkPointParameters}, it changes by about $6\%$ when $\sigma_\ir$ is varied from $10^{-16}$ to $10^{-32}$). The dominant variation in the Yukawa coupling then arises from $\tilde{\lambda}_u$.

In order to reproduce the observed quark masses and mixings, we need to match the effective 5D Yukawa coupling $\tilde{\lambda}(\sigma_\ir)$ evaluated at the minimum of the Goldberger-Wise potential $\sigma_\ir^{\rm min}$ with the values in eq.~(\ref {BenchmarkYukawaCouplings}).
This fixes the combination $\lambda_u k  \, - \, 2 \sqrt{\epsilon} \, \kappa_u  k^2 \, v_\ir$. In order to estimate the size of the remaining, free combination of $\lambda_u$ and $\kappa_u$, we use naive dimensional analysis (NDA) \cite{Manohar:1983md,Chacko:1999hg}. 
Assuming that all loop processes become strong at a cutoff scale $\Lambda$, we write
\be
{\cal L} \, = \, \sqrt{g}\left\{  \frac{\Lambda^5}{\ell_5} {\cal L}_{\rm bulk}+  \delta(y-y_{\rm IR}) \frac{\Lambda^4}{\ell_4} {\cal L}_{\rm brane}    \right\}\, ,
\ee
where $\ell_D = 2^D \pi^{D/2} \Gamma(D/2)$ is the $D$-dimensional loop factor and ${\cal L}_{\rm bulk}$ and ${\cal L}_{\rm brane}$ are functions of the dimensionless ratios $\smash{\partial_A/\Lambda, \, \mathcal{Q}/\Lambda^2 , \, \mathcal{U}/\Lambda^2\, ,\phi/\Lambda^{3/2}}$ and $\smash{\tilde{H}/\Lambda}$. After canonical normalisation of the fields, this gives
\be 
\label{NDAestimate}
\lambda_u k \, = \, d_\lambda \, \frac{\ell_5^{2/3}}{\ell_4^{1/2}} \frac{k}{M_5}\, , \qquad \qquad  \kappa_u k^2 v_\ir  \, = \, d_\kappa \, \frac{\ell_5^{1/3}}{\ell_4^{1/2}} \frac{k^2 v_\ir}{M_5^{7/2}} \, ,
\ee 
where we have used that $\Lambda \sim M_5 \ell_5^{1/3}$ from NDA and the coefficients $d_\lambda$ and $d_\kappa$ are of order one. We next need to estimate the allowed sizes of $k /M_5$ and $v_\ir/M_5^{3/2}$. The AdS curvature scale $k$ is limited by the requirement that higher-curvature terms in the action can be neglected so that the solution to the Einstein equation can be trusted. Using NDA, this gives $k/M_5 \lesssim (3 \pi^3)^{1/3}/5^{1/2}$ \cite{Agashe:2007zd}. Similarly, the VEV $v_\ir$ at the IR brane is limited by demanding that the backreaction of the Goldberger-Wise scalar on the geometry can be neglected. Since we want to ensure this also away from the minimum of the Goldberger-Wise potential, the resulting condition is somewhat more stringent than usual. Indeed, for $\sigma_\ir \ll \sigma_\ir^{\rm min}$ the VEV is well approximated by $\langle \phi \rangle \approx v_\ir \sigma_\ir^{4+\epsilon} e^{(4+\epsilon)k y}$ in the IR. The contribution to the energy-momentum tensor from the kinetic term is then not suppressed by $\epsilon$ (contrary to the case $\sigma_\ir = \sigma_\ir^{\rm min}$). In particular, near the IR brane we have
\be 
T^{MN}_{\phi,\ir} \, \approx \, 8 \, k^2  v_\ir^2 g^{MN}\, .
\ee
Demanding that this is negligible compared to the contribution from the bulk cosmological constant, $T^{MN}_{\rm c.c.} = - 24 M_5^3 k^2 g^{MN}$, gives $v_\ir/M_5^{3/2}  \lesssim \sqrt{3}$. 

\begin{figure}[t]
\centering
\includegraphics[width=12cm]{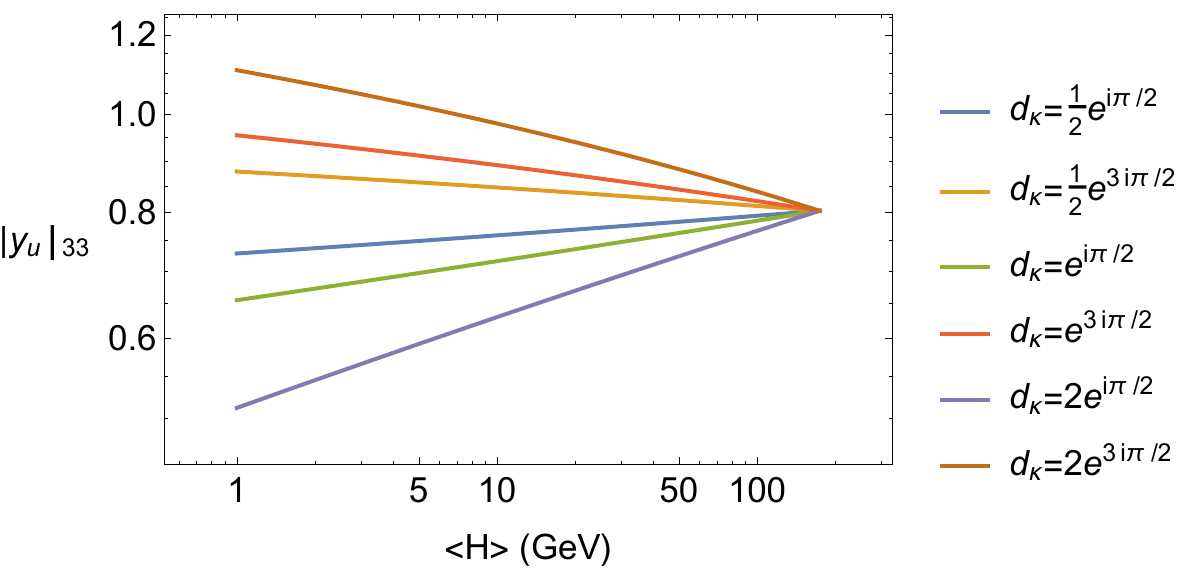}
\caption{\label{YukawaIRbraneCoupling} \small  The top Yukawa coupling eq.~(\ref{4dYukawa}) with $\lambda_u \rightarrow \tilde{\lambda}_u$ given by eq.~(\ref{5dYukawa}), as a function of the Higgs VEV for $\epsilon=1/20$ and different values of $d_\kappa$.}
\end{figure}

For definiteness, we set $k/M_5=1/2$ and $v_\ir/M_5^{3/2}=1$. We then fix $d_\lambda$ for given $d_\kappa$ and $\epsilon$ by the requirement that the 5D Yukawa coupling for the top in eq.~\eqref{BenchmarkYukawaCouplings} is reproduced. We also trade the radion VEV for the Higgs VEV via the relation in eq.~\eqref{RelationHiggsVEVRadionVEV}.
In fig.~\ref{YukawaIRbraneCoupling}, we plot the top Yukawa coupling as a function of the Higgs VEV for $\epsilon=1/20$ and different values of $d_\kappa$ (for all these values $|d_\lambda| \sim 0.3$).  As one can see, the coupling varies with decreasing Higgs VEV. This corresponds to the fact that the derivative of the Goldberger-Wise scalar at the IR brane and its contribution to the Yukawa coupling changes when the IR brane is sent to infinity. In the limit $\{ \sigma_\ir, \langle H \rangle \} \rightarrow  0$, the top Yukawa coupling becomes $|y_u|_{33} \simeq 0.5, 1.1, 0.3, 1.5, 0.7, 2.2$ for $d_\kappa = \frac{1}{2} e^{i \pi/2 }, \frac{1}{2} e^{3 i \pi/2 },  e^{i \pi/2 }, e^{3 i \pi/2 },2 e^{i \pi/2 }, 2 e^{3 i \pi/2 }$ which is still in the perturbative regime.

In summary, this simple construction allows for Yukawa coupling variation of order one during the EW phase transition. When applied to the top quark, it can therefore provide sufficient $CP$-violation for EW baryogenesis.
As discussed in sec.~\ref{sec:flavourconstraints}, implications of this model for flavour and $CP$-violating observables are rather minor.
We next move to what we consider to be the most interesting aspects of our study.

\section{Model II: Large Yukawa couplings from modified fermion profiles}
\label{sec:ModelII}

As reviewed in sec.~\ref{sec:FlavourinRS}, the massless modes of bulk fermions with constant mass terms have profiles along the extra dimension which are localized towards either the UV or IR brane. For our second model, we consider a Yukawa coupling of the Goldberger-Wise scalar to the bulk fermions, giving rise to position-dependent mass terms for the fermions. These modify the profiles of the massless modes and allow for profiles which are localized in the UV and thus decay towards the IR but then `turn around' at some point along the extra dimension and start growing again towards the IR.
Fermions which are UV-localized if the IR brane is at the minimum of the Goldberger-Wise potential can then become IR-localized when the IR brane is moved to infinity. This increases the Yukawa couplings to the Higgs on the IR brane.

The fermionic action in the bulk  is the same as eq.~(\ref{fermionaction})
except for the last term which we replace by 
\be
\label{BulkYukawaCoupling}
c_\mathcal{Q}  \, k \, \overline{\mathcal{Q}} \mathcal{Q} \, \rightarrow  \, \rho_\mathcal{Q} \, \phi \, \overline{\mathcal{Q}} \mathcal{Q} \, ,
\ee
where $\rho_\mathcal{Q}$ has dimension $-1/2$. We consider a similar coupling for the right-handed up-type quarks $\mathcal{U}$. Note that we can again perform unitary transformations such that the kinetic terms and the new Yukawa couplings are diagonal in flavour space. The calculations will be performed in this basis here and below. Furthermore, note that we have assumed that any constant contributions $c k$ to the bulk masses are negligible. 
We expect that, even if they are sizeable, our picture does not change qualitatively. Indeed below we study a Goldberger-Wise scalar with a constant contribution to the VEV. The more general case with separate constant and $y$-dependent contributions to the bulk mass would require a $y$-dependent diagonalization of the action. But we expect that the resulting diagonal bulk masses would then give similar wavefunctions as for the Goldberger-Wise scalar with the constant and $y$-dependent contributions to the VEV. Nevertheless we leave a detailed study of the more general case to future work.
In sec.~\ref{sec:GWprofile}, we work out the consequences of the above coupling for a Goldberger-Wise scalar with a profile as discussed in sec.~\ref{sec:ReviewGW}. In sec.~\ref{sec:modifiedGW}, we then consider a modified profile for the Goldberger-Wise scalar with the aforementioned constant contribution which leads to faster growing Yukawa couplings to the Higgs.

\subsection{Using the Goldberger-Wise scalar}
\label{sec:GWprofile}
The profile of the Goldberger-Wise scalar in eq.~\eqref{vevprofile} has two pieces, $A \, e^{(4+\epsilon) k y}$ and $B \, e^{-\epsilon k y}$. 
As can be seen in fig.~\ref{fig:GWprofile}, the first piece becomes important only close to the IR brane. In order to simplify the calculation, we therefore approximate the profile by the second piece:\footnote{This profile also arises for a vanishing potential on the IR brane, $\lambda_\ir =0$ (though such a scalar no longer stabilizes the extra dimension). Indeed the boundary conditions eqs.~\eqref{boundaryconditionUV} and \eqref{boundaryconditionIR} in this case give 
$A \simeq \frac{\epsilon}{4} v_\uv  \sigma_\ir^{4+2 \epsilon}$ and $B  \simeq  v_\uv$ in the limit of large $\lambda_\uv$.  
Comparing the resulting sizes of the two contributions to the profile, $A \, e^{(4+\epsilon) k y}$ and $B \, e^{-\epsilon k y}$, we find that it is everywhere well approximated by eq.~\eqref{simplescalarvev}.}
\be
\label{simplescalarvev}
\langle \phi \rangle \, \simeq \, v_\uv \, e^{- \epsilon k y} \, .
\ee
Later we will check explicitly that this gives an excellent approximation to using the exact profile in eq.~\eqref{vevprofile}.
The bulk equation of motion for $\mathcal{Q}$ is given by eq.~\eqref{fermioneom} with 
\be 
\label{localc}
c_\mathcal{Q}(y) \, = \, c^{\rm loc}_\mathcal{Q}(y) \, \equiv \, \rho_\mathcal{Q} \langle \phi \rangle / k \, = \, \tilde{c}_\mathcal{Q}  \,  e^{-\epsilon k y} \, ,
\ee
where the constants 
\be
\tilde{c}_\mathcal{Q} \equiv \rho_\mathcal{Q} v_\uv/k
\ee
are dimensionless. The wavefunctions of the left-handed massless modes of $\mathcal{Q}$ then are
\be 
\label{righthandedmasslessmode2}
f_L^{(0)}(y)\, = \, \mathcal{N}_{\tilde{c}_\mathcal{Q}}^{(0)} \, e^{\frac{\tilde{c}_\mathcal{Q}}{\epsilon}e^{- \epsilon k y}} \, ,
\ee
with the modified normalisation constant 
\be 
\label{normalizationmasslessmode2}
\mathcal{N}^{(0)}_{\tilde{c}_\mathcal{Q}} \, = \, \sqrt{\epsilon} \, \left[ \sigma_{_{\rm IR}}^{-1} \, E_{1+\frac{1}{\epsilon}}\hspace*{-.1cm} \left(\frac{-2 \, \tilde{c}_\mathcal{Q} \, \sigma_{_{\rm IR}}^{\epsilon}}{\epsilon}\right) \, - \, E_{1+\frac{1}{\epsilon}}\hspace*{-.1cm} \left(\frac{-2 \, \tilde{c}_\mathcal{Q}}{\epsilon}\right)\right]^{-1/2} 
\ee
and $E_n(x)$ is the exponential integral function. 
For the bulk fermions $\mathcal{U}$ with right-handed massless modes, we redefine $\tilde{c} \rightarrow -\tilde{c}$. Their wavefunctions are then given by eqs.~\eqref{righthandedmasslessmode2} and \eqref{normalizationmasslessmode2} with $\tilde{c}_\mathcal{Q}$ replaced by $\tilde{c}_\mathcal{U}\equiv \rho_\mathcal{U}v_\uv/k$.
\begin{figure}[t]
\centering
\includegraphics[width=5.236cm]{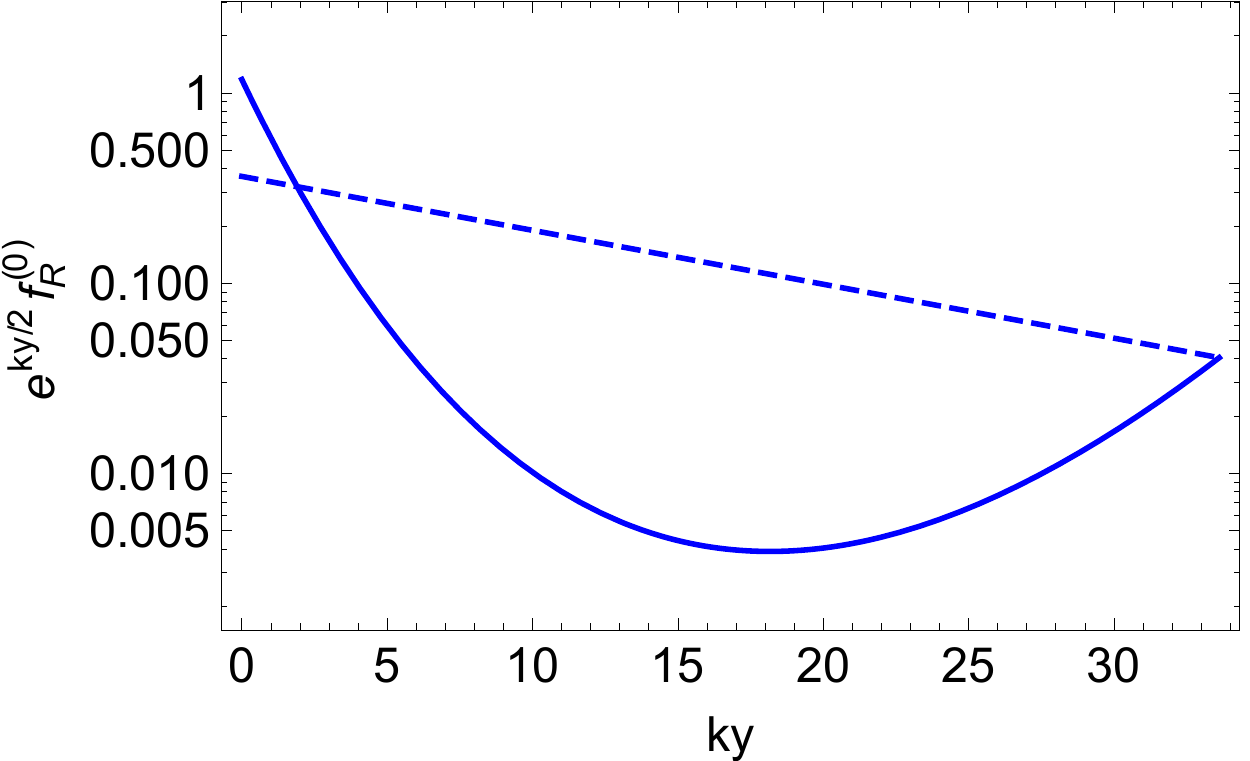}
\includegraphics[width=5.36cm]{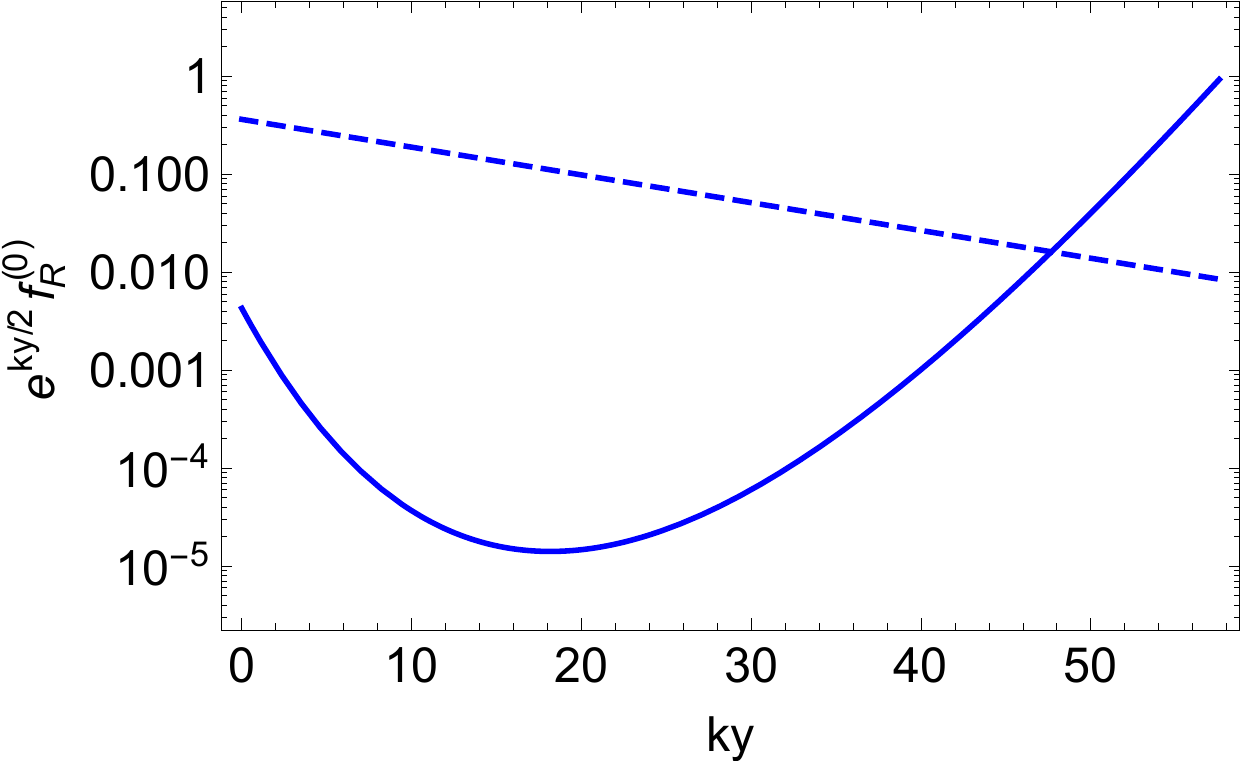}
\includegraphics[width=5.36cm]{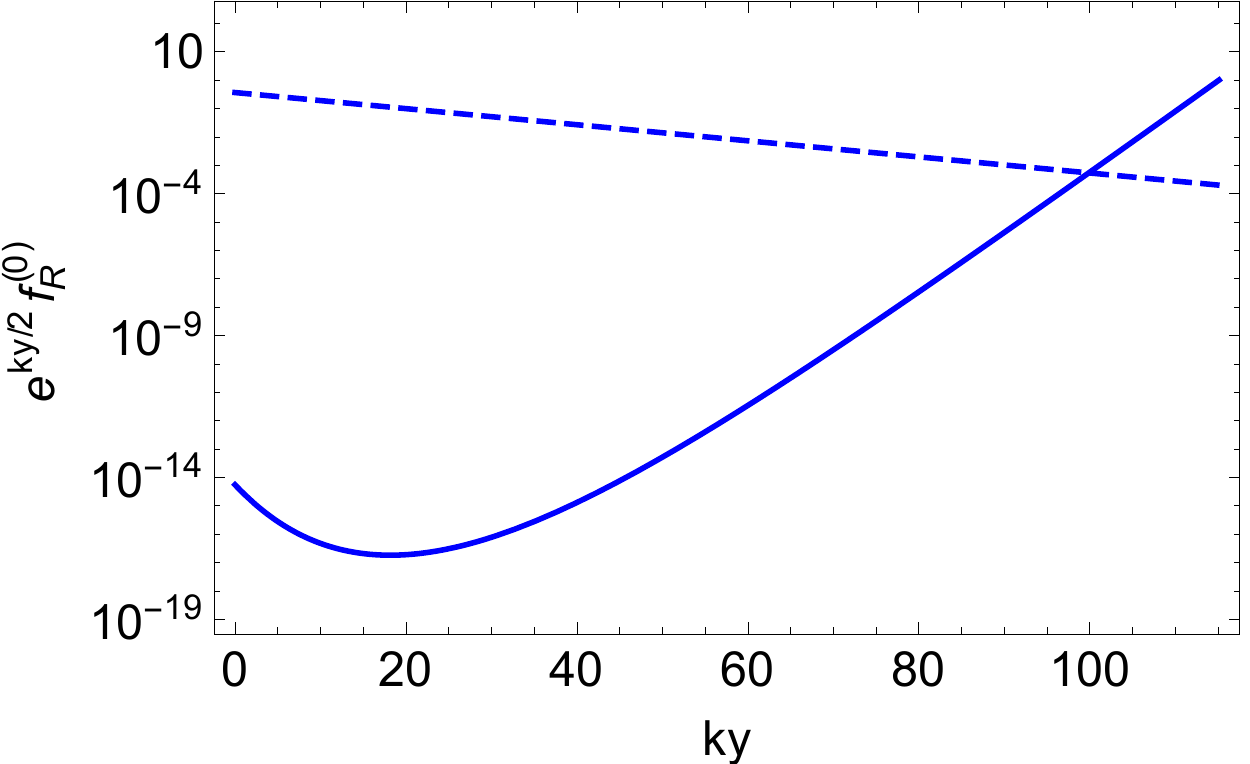}
\includegraphics[width=5.36cm]{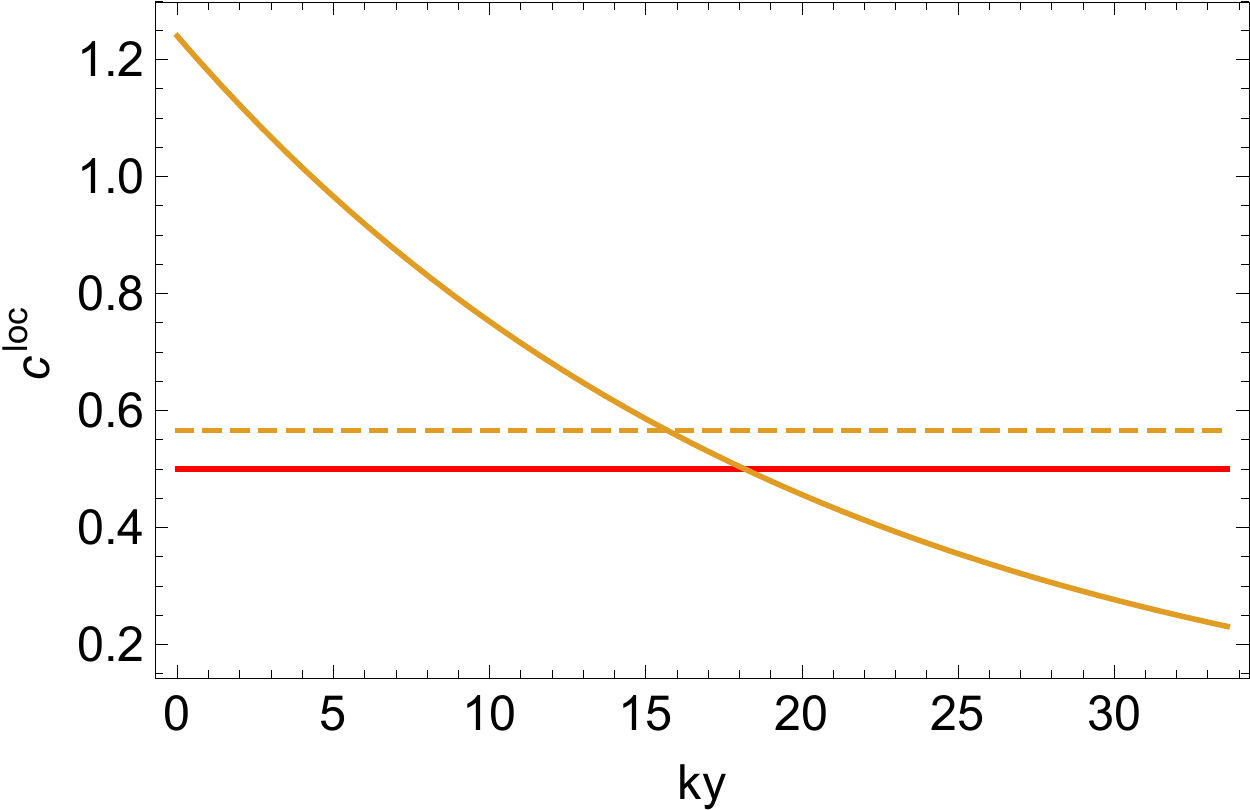}
\includegraphics[width=5.36cm]{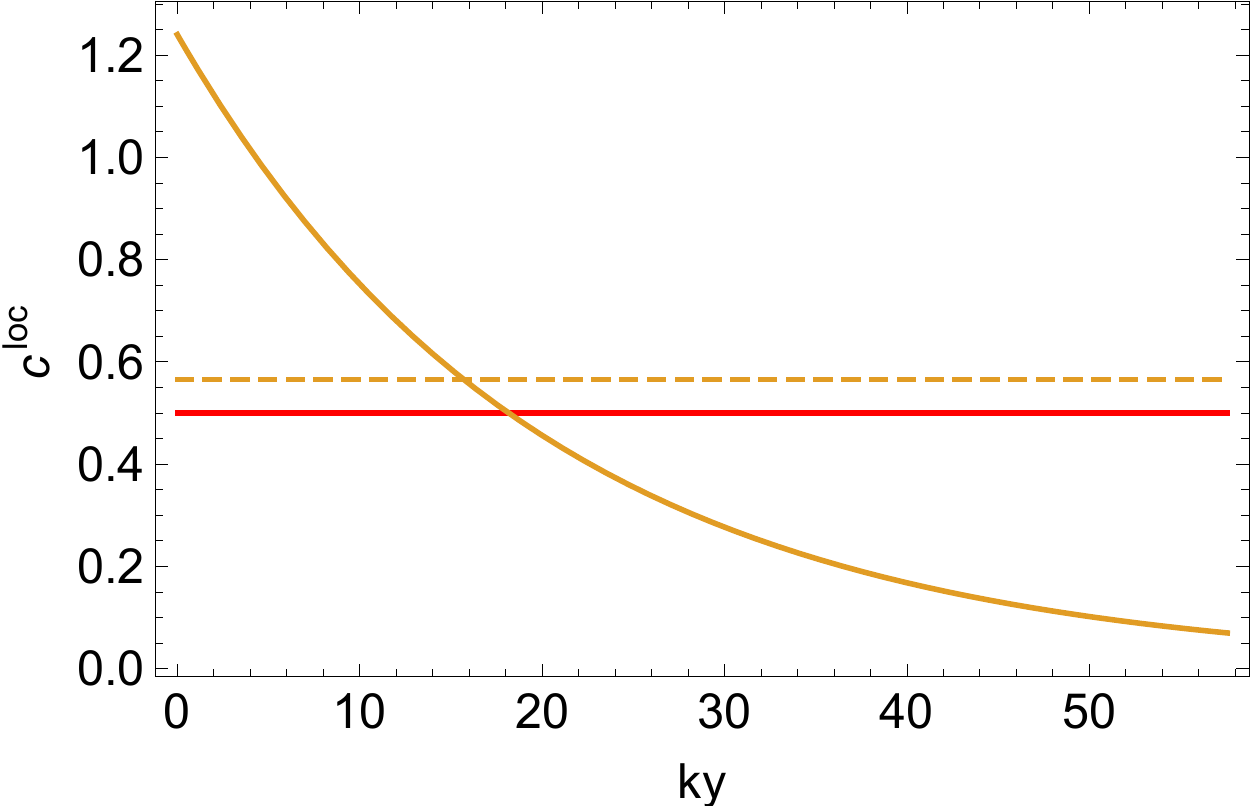}
\includegraphics[width=5.36cm]{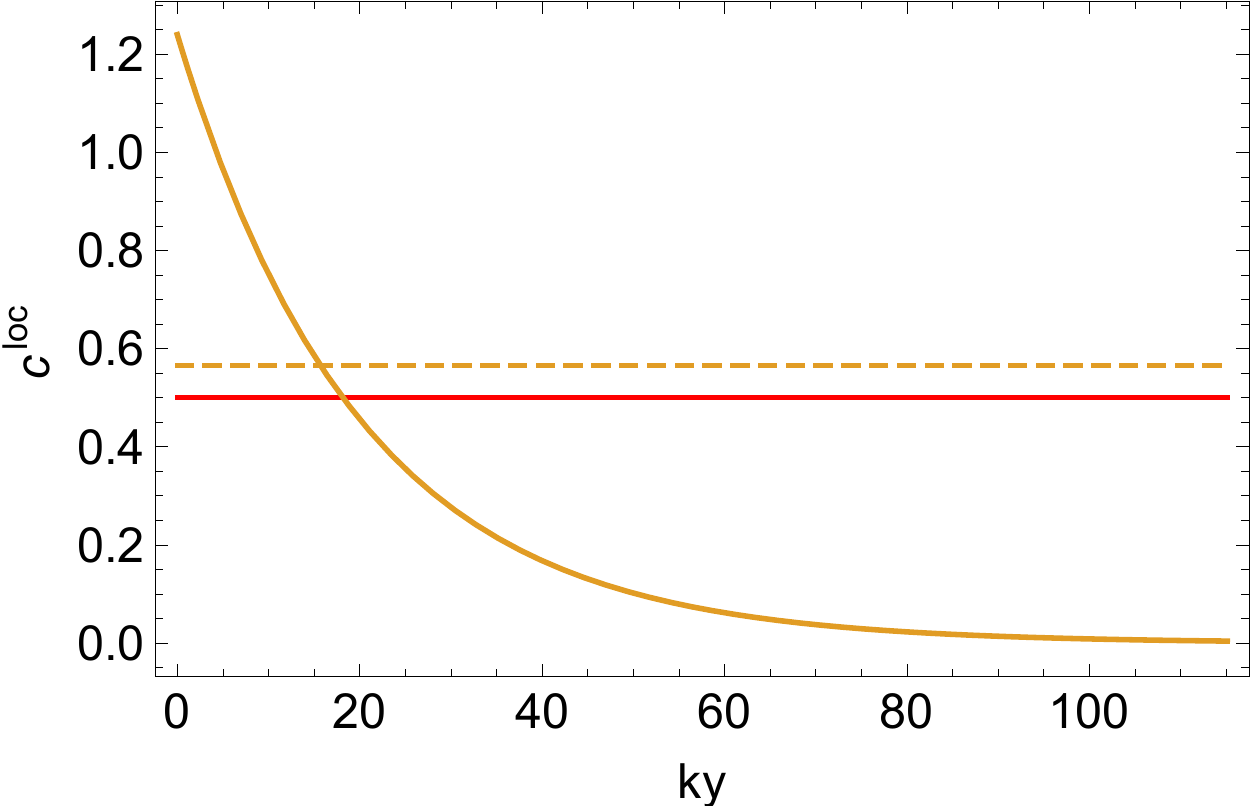}
\caption{\label{fig:wavefunctionsandcloc} \small From left to right, the IR brane is being pushed away from the UV brane with the hierarchies $\sigma_\ir = 2.5 \times 10^{-15}, 10^{-25}$ and $10^{-50}$ respectively. 
Upper panel: The normalized wavefunction of the right-handed charm along the extra dimension.
The solid curve is the wavefunction for the position-dependent bulk mass in eq.~\eqref{localc}, whereas the dashed curve is for the usual case with constant bulk mass. Lower panel: The bulk-mass parameter $c^{\rm loc}$ of the right-handed charm along the extra dimension. The solid curve is again for the position-dependent case in eq.~\eqref{localc} and the dashed curve for the usual constant case. The red curve marks the value $c^{\rm loc}=1/2$ for which the wavefunction changes from decaying to growing towards the IR.}
\end{figure}

In order to fix the parameters $\tilde{c}$, we again use the benchmark point from ref.~\cite{Casagrande:2008hr}.
By demanding that the wavefunction overlap with the IR brane of our fermion profiles agree with that for the fermion profiles with constant bulk mass terms, we can translate their values for $c$ to values for our $\tilde{c}$. Choosing $\epsilon=1/20$ and the hierarchy in the minimum of the radion potential as $\sigma_\ir^{\rm min}=2.5 \times 10^{-15}$, we find for the top-charm sector:
\be
\label{ctvalues}
\tilde{c}_{\mathcal{Q}_2} \, = \, 1.17 \qquad \quad \tilde{c}_{\mathcal{U}_2} \, = \, 1.24 \qquad \quad \tilde{c}_{\mathcal{Q}_3} \, = \, 1.01 \qquad \quad \tilde{c}_{\mathcal{U}_3} \, = \, -1.77  \, .
\ee

\begin{figure}[t]
\centering
\includegraphics[width=8.1cm]{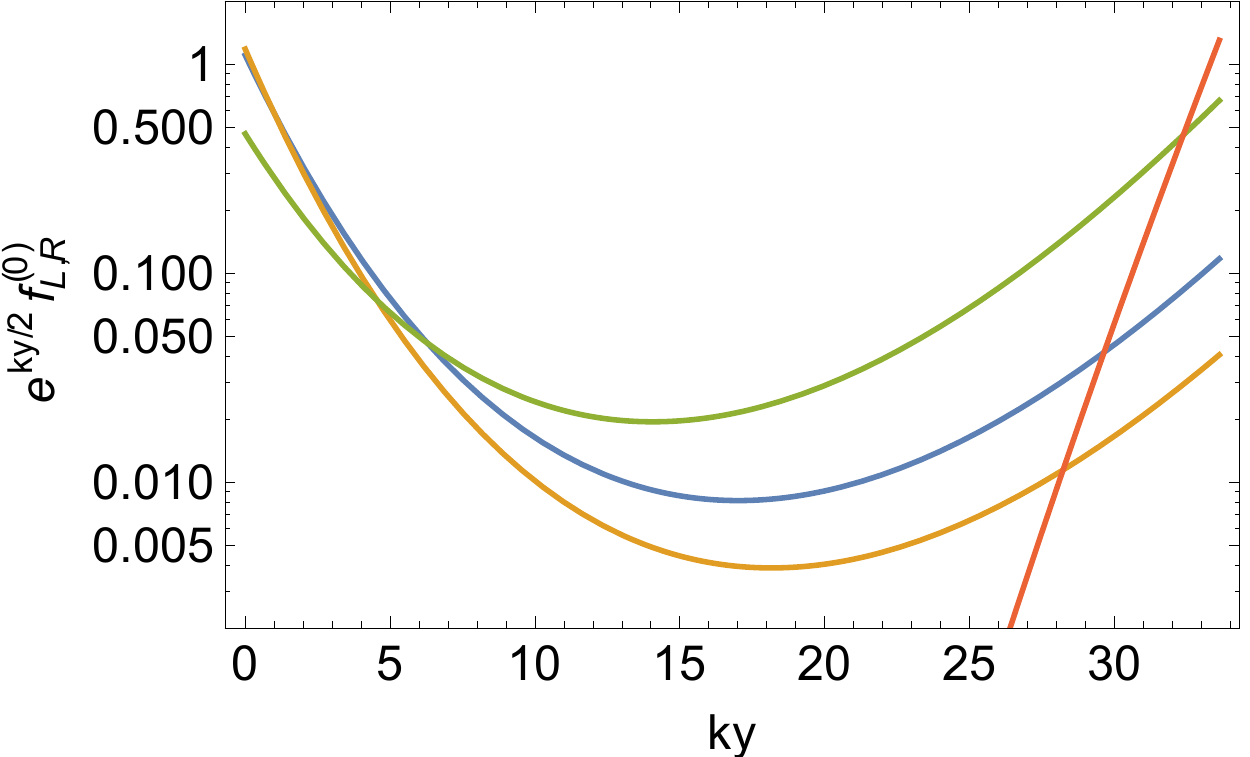}
\includegraphics[width=8.1cm]{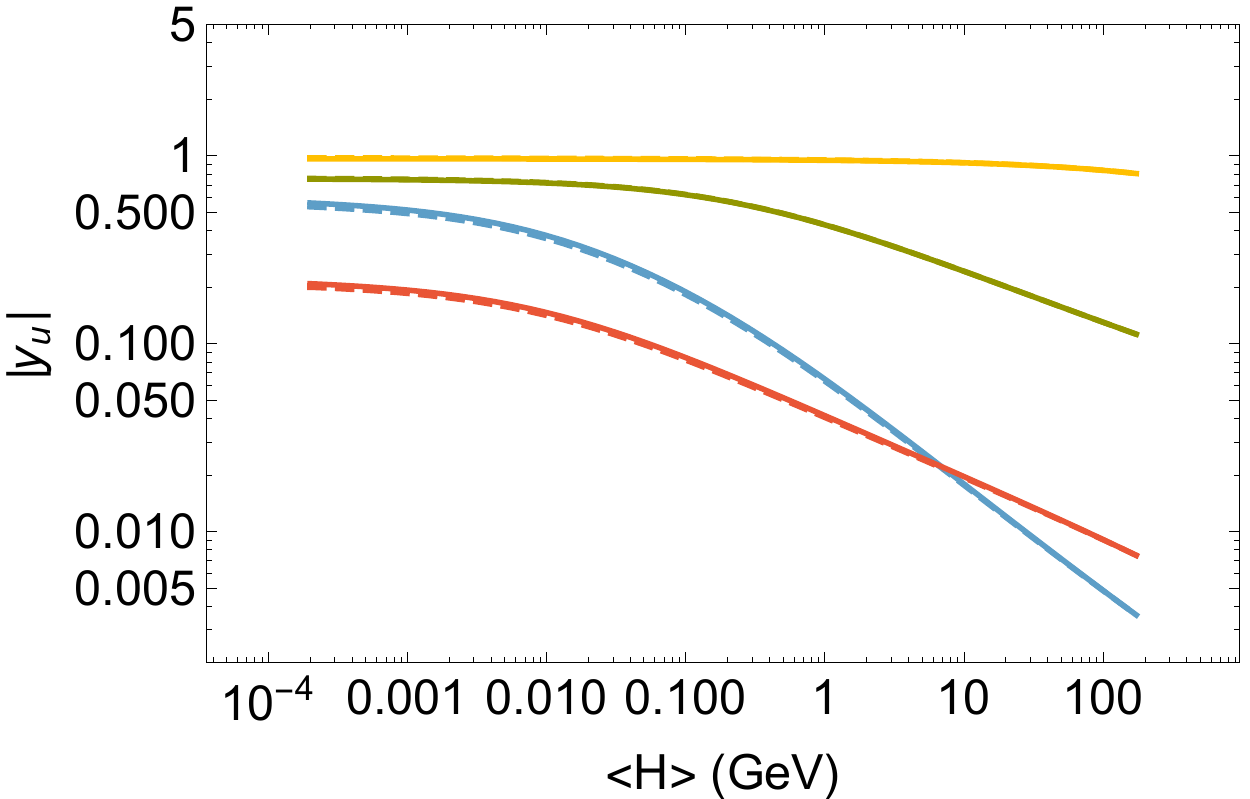}
\caption{\label{fig:wavefunction}
\small (a) Profile along the extra dimension of the left- and right-handed charm (blue and yellow), and the left- and right-handed top (green and red) for the approximation \eqref{simplescalarvev} to the Goldberger-Wise profile. (b) Yukawa couplings $|y_{u}|_{22}$ of the charm (blue), $|y_{u}|_{33}$ of the top (yellow) and the off-diagonal Yukawa couplings $|y_{u}|_{23}$ (green) and $|y_{u}|_{32}$ (red). The solid curves were generated using the approximation \eqref{simplescalarvev} to the Goldberger-Wise profile, whereas for the dashed curves the exact expression \eqref{vevprofile} was used.}
\end{figure}

In the upper panel of fig.~\ref{fig:wavefunctionsandcloc}, we show the resulting wavefunction of the right-handed charm along the extra dimension (mulitplied by $e^{ky/2}$ as this gives the function whose square is normalized to one, cf.~eq.~\eqref{orthonormalityconditions}). The three figures correspond to the hierarchies $\sigma_\ir = 2.5 \times 10^{-15}, 10^{-25}$ and $10^{-50}$ so the sequence from left to right can be understood as going along the bubble wall profile where the IR brane is moved to infinity. As one can see, the wavefunction initially decays when going from the UV to the IR but then starts to grow again. This can be understood as follows: As reviewed in sec.~\ref{sec:FlavourinRS}, for a fermion with constant bulk mass $c k$, the massless mode is UV (IR) localized for $c > 1/2 \; (c<1/2)$. In our setup, the bulk mass is $\rho \langle \phi \rangle = c^{\rm loc} k$ and depends on the position along the extra dimension. In the lower panel of fig.~\ref{fig:wavefunctionsandcloc}, we plot the bulk-mass parameter $c^{\rm loc}$ for the right-handed charm along the extra dimension. The three figures again correspond to the hierarchies $\sigma_\ir = 2.5 \times 10^{-15}, 10^{-25}$ and $10^{-50}$. 
Notice that $c^{\rm loc}>1/2$ near the UV brane and the wavefunction thus decays towards the IR in that region. This changes to $c^{\rm loc}<1/2$ near the IR brane, on the other hand, leading to a growing wavefunction towards the IR. Since $c^{\rm loc}$ is always smaller than $1/2$ sufficiently deep in the IR, we see that in our model all fermions eventually become IR-localized if the IR brane is moved to infinity. This is visible in the upper right plot in fig.~\ref{fig:wavefunctionsandcloc}.
In fig.~\ref{fig:wavefunction}(a), we show all the wavefunctions from the charm-top sector for the case that the radion is at the minimum of its potential, $\sigma_\ir = \sigma_\ir^{\rm min} = 2.5 \times 10^{-15}$.
Note that for the right-handed top, $c^{\rm loc}<1/2$ everywhere and the wavefunction is thus completely localized towards the IR.

As before, we assume that the Higgs is localized on the IR brane. In order to simplify the discussion, we do not couple the Goldberger-Wise scalar to the Yukawa operator on the IR brane as in sec.~\ref{sec:ModelI}. Both effects -- from the coupling in the bulk and on the IR brane -- could of course be present simultaneously and would then give even stronger $CP$-violation during the phase transition.  
The 5D Yukawa coupling of the bulk fermions $\mathcal{Q}$ and $\mathcal{U}$ to the Higgs $\tilde{H}$ on the IR brane is then 
given by eq.~(\ref{localizedYukawa}), leading to the 4D Yukawa coupling in eq.~\eqref{Action4dYukawa} with
\be
\label{ModifiedYukawaCouplings}
y_{u}(\sigma_\ir)  \, = \, \lambda_{u} k \; \mathcal{N}^{(0)}_{\tilde{c}_{\mathcal{Q}}} \mathcal{N}^{(0)}_{\tilde{c}_{\mathcal{U}}} \, \sigma_\ir^{-1} \,  e^{\frac{(\tilde{c}_{\mathcal{Q}} + \tilde{c}_{\mathcal{U}}) \, \sigma_\ir^{\epsilon}}{\epsilon}} \, .
\ee

Let us study the above expression in some limits. Since $\epsilon \ll 1$, the exponential integral functions in the normalization constants are well approximated by the leading term in the expansion 
\be 
\label{EnApprox}
E_n(x) \, = \, \frac{e^{-x}}{x+n}\, \left(1 \, + \, \frac{n}{(n+x)^2} \, + \, \dots \right) 
\ee
for large argument $n$ \cite{AbramowitzStegun}. The expression for the 4D Yukawa coupling then simplifies to
\be 
y_{u}(\sigma_\ir)  \, \approx \, \lambda_{u} k \, \frac{\sqrt{1- 2 \tilde{c}_\mathcal{Q} \sigma_\ir^\epsilon}}{\sqrt{1- \frac{1- 2 \tilde{c}_\mathcal{Q} \sigma_\ir^\epsilon}{1- 2 \tilde{c}_\mathcal{Q} } \, \sigma_\ir e^{2 \tilde{c}_\mathcal{Q} (1- \sigma_\ir^\epsilon)/\epsilon} }} \frac{\sqrt{1- 2 \tilde{c}_\mathcal{U}\sigma_\ir^\epsilon}}{\sqrt{1- \frac{1- 2 \tilde{c}_\mathcal{U}\sigma_\ir^\epsilon}{1- 2 \tilde{c}_\mathcal{U}} \, \sigma_\ir e^{2 \tilde{c}_\mathcal{U}(1- \sigma_\ir^\epsilon)/\epsilon} }} \, .
\ee
From this we see immediately that in the limit $\sigma_\ir \rightarrow 0$ we have
\be 
\label{YukawaIRapprox}
y_{u}(\sigma_\ir)  \, \underset{\sigma_\ir \rightarrow \, 0}{\approx}  \, \lambda_{u} k \, .
\ee
This just reflects the fact that all fermions become IR-localized for $\sigma_\ir \rightarrow 0$ as noted above so that there is no wavefunction suppression of the Yukawa coupling any more. In the limit $\epsilon \rightarrow 0$ we find
\be
y_{u}(\sigma_\ir)  \, \underset{\epsilon \rightarrow 0}{=} \, \lambda_{u} k \,\sqrt{\frac{1-2 \tilde{c}_\mathcal{Q}}{1-\sigma_\ir^{1-2\tilde{c}_\mathcal{Q}}}} \, \sqrt{\frac{1-2 \tilde{c}_\mathcal{U}}{1-\sigma_\ir^{1-2\tilde{c}_\mathcal{U}}}}  \, .
\ee
This agrees with the expression in eq.~\eqref{4dYukawa} for the 4D Yukawa coupling for fermions with constant bulk masses, as is expected since $c^{\rm loc}$ becomes constant for $\epsilon \rightarrow 0$. Similarly, the profile of the massless mode \eqref{righthandedmasslessmode2} agrees with the  profile \eqref{righthandedmasslessmode} for the case of constant bulk masses in that limit.

For a fermion that is localized towards the UV brane, the normalization constant \eqref{normalizationmasslessmode2} depends only weakly on the position $\sigma_\ir$ of the IR brane. We can then neglect the corresponding part in the expression. If both $\mathcal{Q}$ and $\mathcal{U}$ are UV-localized, this gives
\be 
\label{YukawaUVapprox}
y_{u}(\sigma_\ir)  \, \underset{\text{UV loc.}}{\approx}  \, \lambda_{u} k \, \sqrt{2 \tilde{c}_\mathcal{Q}-1} \, \sqrt{2 \tilde{c}_\mathcal{U}-1} \, \sigma_\ir^{-1} \, e^{-\frac{ (\tilde{c}_\mathcal{Q}+\tilde{c}_\mathcal{U}) (1- \sigma_\ir^\epsilon)}{\epsilon}} \, .
\ee
We see that for $\tilde{c}_\mathcal{Q}+\tilde{c}_\mathcal{U}>0$, the exponential decreases if $\sigma_\ir$ becomes smaller. For a certain range of $\sigma_\ir$, this can offset the increase due to the factor of $\sigma_\ir^{-1}$. However, eventually the latter effect starts to dominate and the Yukawa coupling keeps growing with decreasing $\sigma_\ir$. This change happens near a position of the IR brane $\sigma_\ir$ where the wavefunctions turn from decaying to growing towards the IR. For very small $\sigma_\ir$, the approximation leading to eq.~\eqref{YukawaUVapprox} then eventually breaks down because the fields become localized in the IR and the Yukawa coupling is better approximated by eq.~\eqref{YukawaIRapprox}.

In fig.~\ref{fig:wavefunction}(b), we plot the Yukawa couplings for the top-charm sector using the parameters for the benchmark point in eqs.~\eqref{BenchmarkYukawaCouplings} and \eqref{ctvalues}. We again trade the radion VEV for the Higgs VEV via the relation in eq.~\eqref{RelationHiggsVEVRadionVEV}. We see that the Yukawa couplings grow with decreasing Higgs VEV (or radion VEV). In particular, the charm coupling $|y_u|_{22}$ and the charm-top coupling $|y_u|_{23}$ become of order 1  for Higgs VEVs less than about $10^{-2}$ GeV. On the other hand, the top coupling $|y_u|_{33}$ remains almost constant. This is due to the fact that the right-handed top is highly localized in the IR for any position of the IR brane (cf.~fig.~\ref{fig:wavefunction}(a)).

So far we have approximated the Goldberger-Wise scalar by the simplified profile in eq.~\eqref{simplescalarvev}. Let us now consider the exact profile in eq.~\eqref{vevprofile}. For the left-handed massless modes of $\mathcal{Q}$, the equation of motion \eqref{fermioneom} is then solved by
\be 
\label{righthandedmasslessmodeGW}
f_L^{(0)}(y)\, = \, \mathcal{N}_{\tilde{c}_\mathcal{Q}}^{(0)} \, e^{\frac{\tilde{c}_\mathcal{Q} }{\epsilon}e^{- \epsilon k y}-\frac{\tilde{c}_\mathcal{Q}\,\tilde{\sigma} }{4+\epsilon} e^{(4+\epsilon) k y}} \, ,
\ee
where again $\tilde{c}_\mathcal{Q} \equiv \rho_\mathcal{Q} \, v_\uv/k$ and $\tilde{\sigma} \equiv \sigma_\ir^{4+2 \epsilon} ((\sigma_\ir^{\rm min}/\sigma_\ir)^\epsilon -1 )$.
The normalization constant $ \mathcal{N}_{\tilde{c}_\mathcal{Q}}^{(0)}$ does not allow for an analytic expression and needs to be evaluated numerically from the orthonormality condition \eqref{orthonormalityconditions}.
As before, we redefine $\tilde{c} \rightarrow -\tilde{c}$ for the bulk fermions $\mathcal{U}$ with right-handed massless modes so that their wavefunctions are given by eq.~\eqref{righthandedmasslessmodeGW} with $\tilde{c}_\mathcal{Q} \rightarrow \tilde{c}_\mathcal{U}$.

In fig.~\ref{fig:wavefunction}(b), we plot the resulting Yukawa couplings for the benchmark point in eqs.~\eqref{BenchmarkYukawaCouplings} and \eqref{ctvalues} as dashed lines using the same colour code as for the approximate profile \eqref{simplescalarvev}. As one can see, the difference between using the exact and approximate profiles is marginal (the charm coupling $|y_u|_{22}$ and the top-charm coupling $|y_u|_{32}$ differ by about $5\%$ at $10^{-5} \, \text{GeV}$ and it is even less for the other couplings).
This can be understood as follows: 
At the minimum of the Goldberger-Wise potential, for $\sigma_\ir=\sigma_\ir^{\rm min}$, we have $A\simeq -\sqrt{\epsilon/4} \, v_\uv (\sigma_\ir^{\rm min})^{4+2\epsilon}$ as follows from eqs.~\eqref{leadingA} and \eqref{hierarchyrelation}. The profile of the Goldberger-Wise scalar is then everywhere well approximated by the simple profile in eq.~\eqref{simplescalarvev}.
For $\sigma_\ir \ll \sigma_\ir^{\rm min}$, on the other hand, we have $A \simeq v_\uv  \sigma_\ir^{4+\epsilon} (\sigma_\ir^{\rm min})^\epsilon$ and the contribution $A \, e^{(4+\epsilon) k y}$ to the profile becomes potentially important. Comparing with $B \, e^{-\epsilon k y}$, we see that the former dominates over the latter in the region
\be 
\sigma_\ir \, \leq \, e^{- k y} \, \lesssim \, \sigma_\ir \left(\frac{\sigma_\ir^{\rm min}}{\sigma_\ir}\right)^{\frac{\epsilon}{4+2 \epsilon}} \, .
\ee
Even for $\sigma_\ir \sim 10^{-100}$, this is a relatively small region $\sigma_\ir \leq  e^{- k y} \lesssim 10 \, \sigma_\ir$ near the IR brane and the difference between using the exact and the approximate profiles is correspondingly small.

\subsection{Using the Goldberger-Wise scalar with a modified profile}
\label{sec:modifiedGW}

In the last section, we have seen that position-dependent bulk masses for the fermions from the Goldberger-Wise scalar
naturally allow for Yukawa couplings which grow when the IR brane is moved to infinity. However, the Yukawa couplings involving the charm become of order 1 only for relatively small radion or Higgs VEVs as can be seen in fig.~\ref{fig:wavefunction}(b). This means that the coupling and the resulting $CP$-violation is large only in a small fraction of the bubble wall during the phase transition which suppresses the produced baryon asymmetry. In order to improve on this, notice that the local bulk-mass parameter $c^{\rm loc}$ in eq.~\eqref{localc} cannot become negative if it is positive near the UV brane (as is necessary for UV-localized fermions). Since the fermion wavefunctions are more IR-localized the smaller $c$ is, having $c^{\rm loc}$ become negative leads to faster growing wavefunctions and thus Yukawa couplings. On the other hand, in order to allow for UV-localized fermions when the radion is at the minimum of the Goldberger-Wise potential, we need positive $c^{\rm loc}$ near the UV brane. Both requirements can be satisfied if the scalar VEV that gives rise to the bulk masses changes sign between the UV and IR brane. We will now discuss how the Goldberger-Wise scalar can obtain such a VEV using a small modification to the original proposal.
To this end, we consider the action
\be
\label{modGWaction}
S \, \supset \, \int d^5 x \,\sqrt{g} \, \left(\frac{1}{2}\partial_A \phi \, \partial^A \phi  -  \frac{m_\phi^2}{2} (\phi+\beta)^2  -  \delta(y) \, \lambda_\uv \tilde{V}_\uv(\phi)  -  \delta(y-y_\ir) \, \lambda_\ir \tilde{V}_\ir(\phi) \right) \, ,
\ee
where $\beta$ is a constant. We choose the boundary potentials $\tilde{V}_\uv(\phi)$ and $\tilde{V}_\ir(\phi)$ to have minima at respectively $\langle \phi \rangle = v_\uv$ and $\langle \phi \rangle = -v_\ir$ (with definitive signs, as opposed to the boundary potentials in eq.~\eqref{BoundaryPotentials} which are degenerate for field values with positive and negative signs). 
Note also that, up to a constant, a bulk potential with a mass term and a tadpole can always be written in the above form. The tadpole just shifts the VEV by a constant. Indeed defining the shifted field $\tilde{\phi}\equiv \phi + \beta$, the tadpole disappears from the action for $\tilde{\phi}$ and the bulk potential only contains a mass term. The VEV $\langle \tilde{\phi} \rangle$ therefore again has the form in eq.~\eqref{vevprofile}. Going back to the original field, we see that
\be 
\label{modifiedvevprofile}
\langle \phi \rangle \, = \,-\beta \, + \, A \, e^{(4+\epsilon) k y} \, + \, B \, e^{-\epsilon k y} \, ,
\ee
where as before $\smash{\epsilon = \sqrt{4+m_\phi^2/k^2}-2}$. The integration constants $A$ and $B$ are determined by the boundary potentials. In terms of the shifted field $\tilde{\phi}$, the minima of the latter are at $\tilde{v}_\uv\equiv \beta + v_\uv$ and $\tilde{v}_\ir\equiv \beta- v_\ir$. In the limit of large couplings $\lambda_\uv, \lambda_\ir$, the integration constants are therefore given by eqs.~\eqref{leadingA} and \eqref{leadingB} with $v_\uv,v_\ir$ replaced by $\tilde{v}_\uv,\tilde{v}_\ir$. Similarly, the radion potential is again given by eq.~\eqref{RadionPotential}.
Choosing $\tilde{v}_\uv, \tilde{v}_\ir >0$, to leading order in $\epsilon$ the radion is then stabilized at
\be
\sigma_\ir^{\rm min} \, \equiv \, \left(\frac{\tilde{v}_\ir}{\tilde{v}_\uv} \right)^{1/\epsilon} \, .
\ee

As before, we drop the contribution $A \, e^{(4+\epsilon) k y}$ to the scalar profile since it gives only a negligible correction to the wavefunctions and couplings. The bulk equations of motion for $\mathcal{Q}$ are then given by eq.~\eqref{fermioneom} with
\be 
\label{localcmod}
c^{\rm loc}_\mathcal{Q}(y) \, \equiv \,  \rho_\mathcal{Q} \langle \phi \rangle / k \, \simeq \, c_\mathcal{Q}  \, + \, \tilde{c}_\mathcal{Q} \, e^{-\epsilon k y} \, ,
\ee 
where $c_\mathcal{Q} \equiv -\beta \rho_\mathcal{Q}/k$ and $\tilde{c}_\mathcal{Q} \equiv \tilde{v}_\uv \rho_\mathcal{Q}/k$ and similarly for $\mathcal{U}$. For the left-handed massless modes of $\mathcal{Q}$, this yields
\be
\label{righthandedmasslessmodeSC}
f_L^{(0)}(y)\, = \, \mathcal{N}_{\tilde{c}_\mathcal{Q},c_\mathcal{Q}}^{(0)} \, e^{-c_\mathcal{Q} \, k y + \frac{\tilde{c}_\mathcal{Q} }{\epsilon} e^{- \epsilon k y}} 
\ee
with
\be 
\mathcal{N}^{(0)}_{\tilde{c}_\mathcal{Q},c_\mathcal{Q}} \, = \, \sqrt{\epsilon} \, \left[ \sigma_{_{\rm IR}}^{2 c_\mathcal{Q}-1} \, E_{1+\frac{1-2 c_\mathcal{Q}}{\epsilon}}\hspace*{-.1cm} \left(\frac{-2 \, \tilde{c}_\mathcal{Q} \, \sigma_{_{\rm IR}}^{\epsilon}}{\epsilon}\right) \, - \, E_{1+\frac{1-2 c_\mathcal{Q}}{\epsilon}}\hspace*{-.1cm} \left(\frac{-2 \, \tilde{c}_\mathcal{Q}}{\epsilon}\right)\right]^{-1/2}  .
\ee
We redefine $c,\tilde{c} \rightarrow -c,-\tilde{c}$ for the bulk fermions $\mathcal{U}$ with right-handed massless modes so that their wavefunctions are given by the above expression with $c_\mathcal{Q},\tilde{c}_\mathcal{Q} \rightarrow c_\mathcal{U},\tilde{c}_\mathcal{U}$. The 4D Yukawa couplings for the up-type quarks are then given by eq.~\eqref{Action4dYukawa} with
\be
\label{ModifiedYukawaCouplingsMod}
y_{u}(\sigma_\ir)  \, = \, \lambda_{u} k \; \mathcal{N}^{(0)}_{\tilde{c}_{\mathcal{Q}},c_\mathcal{Q}}\, \mathcal{N}^{(0)}_{\tilde{c}_{\mathcal{U}},c_\mathcal{U}} \, \sigma_\ir^{c_\mathcal{Q}+c_\mathcal{U}-1} \,  e^{\frac{(\tilde{c}_{\mathcal{Q}} + \tilde{c}_{\mathcal{U}}) \, \sigma_\ir^{\epsilon}}{\epsilon}} \, .
\ee

\begin{figure}[t]
\centering
\includegraphics[width=8.1cm]{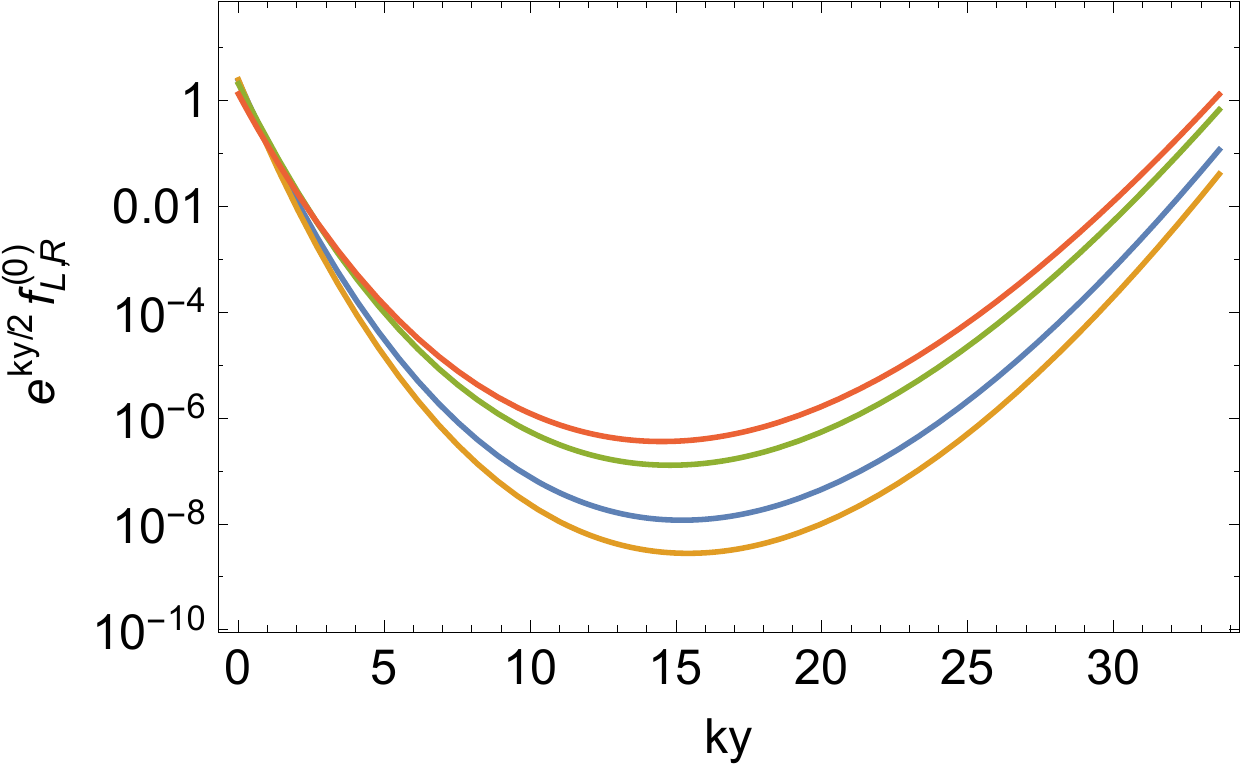}
\includegraphics[width=8.1cm]{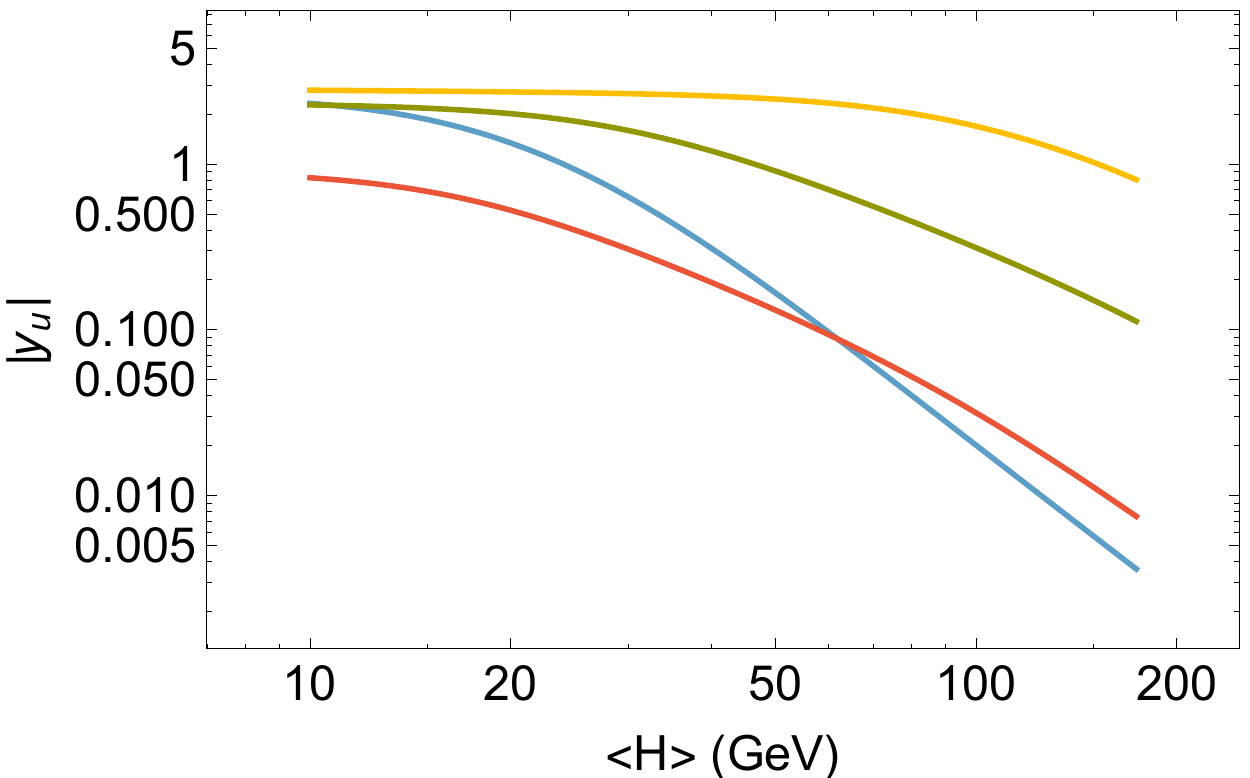}
\caption{ \label{fig:couplingSC} \small (a) Profile along the extra dimension of the left- and right-handed charm (blue and yellow), and the left- and right-handed top (green and red). (b) Yukawa couplings $|y_{u}|_{22}$ of the charm (blue), $|y_{u}|_{33}$ of the top (yellow) and the off-diagonal Yukawa couplings $|y_{u}|_{23}$ (green) and $|y_{u}|_{32}$ (red). }
\end{figure}

Choosing $c<0<\tilde{c}$ and $c+\tilde{c}>0$, the bulk-mass parameter $c^{\rm loc}$ is positive near the UV brane but can become negative in the IR. For definiteness, we set the parameters that determine the Goldberger-Wise potential as\footnote{Together this fixes $\tilde{v}_\ir \simeq (\sigma_\ir^{\rm min})^\epsilon \tilde{v}_\uv \simeq 0.74 \, k^{3/2}$ which in turn requires $v_\ir=\beta-\tilde{v}_\ir \simeq 0.76 \, k^{3/2}$, while $v_\uv=\tilde{v}_\uv-\beta=2.5 k^{3/2}$.} $\beta=1.5 \, k^{3/2}$, $\tilde{v}_\uv=4 \, k^{3/2}$ and as before $\epsilon=1/20$ and $\sigma_\ir^{\rm min}=2.5\cdot 10^{-15}$. We can then choose the couplings $\rho_\mathcal{Q}$ and $\rho_\mathcal{U}$ to achieve different localizations for the zero-mode wavefunctions. Let us again consider the benchmark point from ref.~\cite{Casagrande:2008hr} for which the relevant parameters are given in eqs.~\eqref{BenchmarkYukawaCouplings} and \eqref{BenchmarkPointParameters}. We then demand that the overlaps with the IR brane of our zero-mode wavefunctions at the minimum of the radion potential, $\sigma_\ir=\sigma_\ir^{\rm min}$, reproduce those for the benchmark point. For the top and charm this gives
\be 
\label{rhovalues}
\rho_{\mathcal{Q}_2}  \, = \, 1.35\, k^{-1/2} \qquad \rho_{\mathcal{U}_2}  \, = \, 1.43 \, k^{-1/2} \qquad \rho_{\mathcal{Q}_3}  \, = \, 1.22\, k^{-1/2} \qquad \rho_{\mathcal{U}_3}  \, = \, 1.15 \, k^{-1/2} \, .
\ee
In fig.~\ref{fig:couplingSC}(a), we plot the resulting profiles along the extra dimension. Notice that, compared with the case shown in fig.~\ref{fig:wavefunction}(a), the wavefunctions initially decay much faster towards the IR and then similarly grow much faster beyond the turning point. Correspondingly, we expect that the Yukawa couplings increase more quickly if we move the IR brane to infinity. 
Trading the radion VEV for the Higgs VEV via the relation in eq.~\eqref{RelationHiggsVEVRadionVEV}, we plot the Yukawa couplings as a function of the latter in fig.~\ref{fig:couplingSC}(b).
Comparing with fig.~\ref{fig:wavefunction}(b), we see that indeed the Yukawa couplings grow much faster. In particular, the charm coupling $|y_u|_{22}$ and the charm-top coupling $|y_u|_{23}$ become of order one already for Higgs VEVs around 20 GeV (compared to $10^{-2}$ GeV in the other case).

Such a variation of the Yukawa couplings during the EW phase transition is shown to provide a sufficient source of $CP$-violation to obtain the correct amount of baryon asymmetry during EW baryogenesis, see ref.~\cite{dynamicalyukawas}.

\section{Constraints from flavour- and $CP$-violation}
\label{sec:flavourconstraints}

We will now discuss how flavour- and $CP$-violating processes are modified in our two models compared to the usual scenario in which couplings of the Goldberger-Wise scalar to the bulk fermions are neglected. We remind that in order to obtain sufficient $CP$-violation during the electroweak phase transition it is sufficient to couple the Goldberger-Wise scalar to the top quark in model I or to the top-charm sector in model II. 
However, to be conservative, we will here assume that such couplings exist for all flavours.

We focus on the dominant constraints on the KK scale which arise from $CP$-violation in $K$-$\overline{K}$-mixing \cite{Csaki:2008zd} and from the neutron EDM \cite{Agashe:2004cp}. In secs.~\ref{epsilonK} and \ref{neutronEDM}, we first review the usual contributions to these quantities from SM particles and their higher KK modes. We then discuss modifications that arise in model II due to the position-dependent bulk masses for the fermions. In particular, these lead to decreased overlap integrals of the SM particles with KK gluons and thereby alleviate the contraint from $CP$-violation in $K$-$\overline{K}$-mixing. Since the fermions have the usual constant bulk masses in model I, on the other hand, no such modifications arise in this case. In secs.~\ref{subsec:GWconstraints}, we then consider processes mediated by the Goldberger-Wise scalar which are relevant for both models. 

\begin{figure}[t]
\centering
\includegraphics[width=8cm]{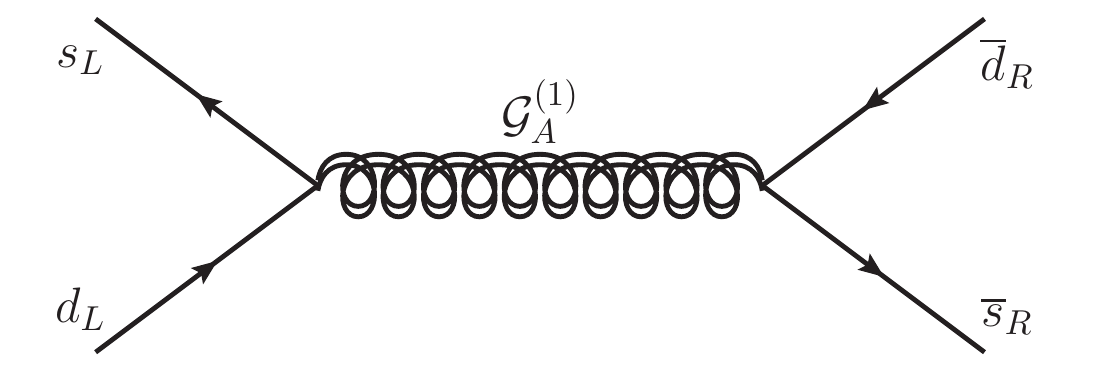}
\includegraphics[width=8cm]{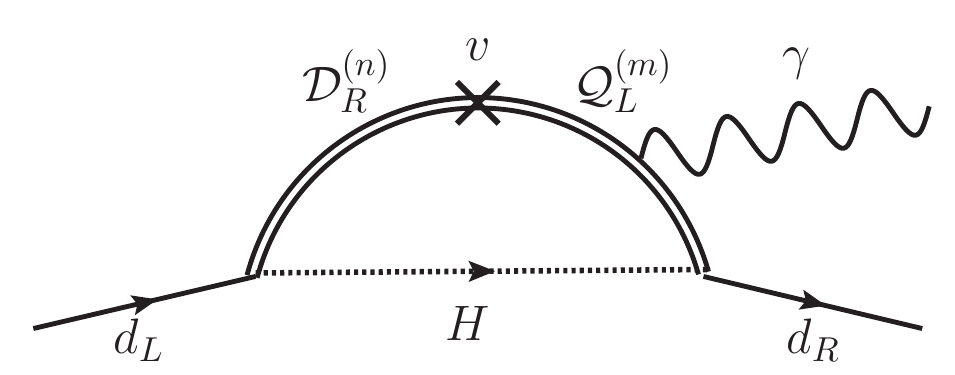}
\caption{\label{fig:FeynmanDiagram} \small The most important, new contribution to (a) $CP$-violation in $K$-$\overline{K}$-mixing and (b) the neutron EDM. Double lines denote KK modes.}
\end{figure}

\subsection{Constraints from the tree-level contribution of KK gluons to $\epsilon_K$}
\label{epsilonK}

An important constraint arises from $\epsilon_K$ which measures $CP$-violation in $K$-$\overline{K}$-mixing.
The most important, new contribution to this quantity is mediated by the first KK mode of the gluon, $K \rightarrow \mathcal{G}^{(1)}_\mu \rightarrow \overline{K}$ \cite{Csaki:2008zd}. The corresponding Feynman diagram is shown in fig.~\ref{fig:FeynmanDiagram}(a). The relevant coupling of the bulk gluon $\mathcal{G}_A$ to the left-handed quark doublets $\mathcal{Q}$ reads
\be
S \, \supset \, \int d^5 x \sqrt{g} \,i \, g^s_5 \,  \mathcal{G}_A \, E^A_a\, \overline{\mathcal{Q}} \, \gamma^a \, \mathcal{Q} \, ,
\ee
where $g^s_5$ is the 5D gauge coupling of QCD. The couplings to the right-handed up-type quarks $\mathcal{U}$ and down-type quarks $\mathcal{D}$ are similar. Expanding the gluon as 
\be
\mathcal{G}_\mu(x,y) = \sqrt{k} \, \sum_n \mathcal{G}_\mu^{(n)}(x) f_\mathcal{G}^{(n)}(y)
\ee
and integrating over the extra dimension, the coupling of the first KK mode of $\mathcal{G}_A$ to the zero-modes of $\mathcal{Q}$ reads
\be 
S \, \supset \, \int d^4 x \, i \,\mathcal{G}^{(1)}_\mu \, \overline{\mathcal{Q}}_L^{(0)} \tilde{g}_{_{\mathcal{Q}L}} \overline{\sigma}^\mu  \, \mathcal{Q}_L^{(0)}  \, ,
\ee
where $\tilde{g}_{_{\mathcal{Q}L}}$ involves an overlap integral over the fermion and gluon wavefunctions. Similar couplings exist for the right-handed quarks. After electroweak symmetry breaking, we rotate the fields $\mathcal{U}^{(0)}_R \rightarrow U_{_{Ru}} u_R$ etc.~in order to obtain diagonal mass matrices. The unitary rotation matrices are hierarchical, with elements
\be 
\left|U_{_{Lu}}\right|_{ij} \, \sim \, \left|U_{_{Ld}}\right|_{ij} \, \sim \, \frac{f_{\mathcal{Q}_i}}{f_{\mathcal{Q}_j}}\, , \qquad \quad  \left|U_{_{Ru}}\right|_{ij} \, \sim \, \frac{f_{\mathcal{U}_i}}{f_{\mathcal{U}_j}}\, , \quad \qquad \left|U_{_{Rd}}\right|_{ij} \, \sim \, \frac{f_{\mathcal{D}_i}}{f_{\mathcal{D}_j}}
\ee
for $i \leq j$, where $f_{\mathcal{Q}_i} \equiv e^{k y_\ir/2} f^{(0)}_{{\mathcal{Q}_iL}}(y_\ir)$ are the wavefunction overlaps of the zero-modes with the IR brane with $f_{\mathcal{Q}_3} > f_{\mathcal{Q}_2}>f_{\mathcal{Q}_1}$ and similarly for $\mathcal{U}$ and $\mathcal{D}$. In terms of the fields with diagonal mass matrices, we then have
\be 
\label{GluonCoupling}
S \, \supset \, \int d^4 x \, i \,\mathcal{G}^{(1)}_\mu\, \overline{u}_L \hat{g}_{_{\mathcal{Q}L}} \overline{\sigma}^\mu  \, u_L  \, ,
\ee
where 
\be 
\hat{g}_{_{\mathcal{Q}L}} \, \sim \, g^s_5 k^{1/2}  
\left( \begin{array}{ccc}
\smash{\alpha + f_{\mathcal{Q}_1}^2 \, (\gamma_{_{\mathcal{Q}_1}}\hspace{-.1cm} +\gamma_{_{\mathcal{Q}_2}}\hspace{-.1cm} +\gamma_{_{\mathcal{Q}_3}})} & f_{\mathcal{Q}_1} f_{\mathcal{Q}_2} \, (\gamma_{_{\mathcal{Q}_2}}\hspace{-.1cm}+\gamma_{_{\mathcal{Q}_3}}) & f_{\mathcal{Q}_1} f_{\mathcal{Q}_3} \, \gamma_{_{\mathcal{Q}_3}} \\
. & \alpha +f_{\mathcal{Q}_2}^2 \, (\gamma_{_{\mathcal{Q}_2}}\hspace{-.1cm} +\gamma_{_{\mathcal{Q}_3}})  & f_{\mathcal{Q}_2} f_{\mathcal{Q}_3} \, \gamma_{_{\mathcal{Q}_3}} \\
. & . &  \alpha +f_{\mathcal{Q}_3}^2 \, \gamma_{_{\mathcal{Q}_3}} 
\end{array} \right) 
\ee
is a symmetric matrix and $u_L$ a vector in flavour space. Furthermore, 
$\smash{\alpha \simeq 1 /\log(m^{(1)}_\mathcal{G}/k)}$ 
with $m^{(1)}_\mathcal{G}$ being the mass of the first gluon KK mode and
\be 
\label{DefinitionGamma}
\gamma_{_{\mathcal{Q}_i}} \, \equiv \, \sqrt{2} \,  k \,\int_0^{y_\ir} dy \,  e^{2k(y-y_\ir)} \frac{J_1\Bigl(\frac{m^{(1)}_\mathcal{G}}{k} e^{ky}\Bigr)}{J_1(m^{(1)}_\mathcal{G}/m_\ir)} \, \left( \frac{f^{(0)}_{\mathcal{Q}_iL}(y)}{ f^{(0)}_{\mathcal{Q}_iL}(y_\ir) }\right)^2 \, ,
\ee
where $m_\ir \equiv \sigma_\ir k$ is the warped-down AdS scale. The couplings to the other quarks $d_L$, $u_R$ and $d_R$ are given by analogous expressions.

The most important constraint on $CP$-violation in $K$-$\overline{K}$-mixing arises from the effective operator
\be 
\label{EffectiveOperator}
\mathcal{L} \, \supset \, -C^4_K \, \overline{d}_R s_L \overline{d}_L s_R\, ,
\ee
where the first and the second pair form color singlets (and $d_{L,R}$ is now the down quark, not a vector in flavour space). This leads to (see e.g.~table 2 in ref.~\cite{Csaki:2008zd}) $\im \, C^4_K < (1.6\cdot 10^5 \, \text{TeV})^{-2}$. Integrating out the first gluon KK mode, we can estimate the Wilson coefficient as
\be 
\label{GluonMediatedC4K}
C^4_K \, \sim \,  \frac{m_d \, m_s}{v_{\rm EW}^2 \, \lambda_*^2}\,\frac{\bigl[g_5^s\bigr]^2 k}{\bigl[m^{(1)}_\mathcal{G}\bigr]^2}\, (\gamma_{_{\mathcal{Q}_2}} \hspace{-.1cm}  + \gamma_{_{\mathcal{Q}_3}}) \, (\gamma_{_{\mathcal{D}_2}} \hspace{-.1cm}  + \gamma_{_{\mathcal{D}_3}}) \, ,
\ee
where $m_d \sim \lambda_* f_{\mathcal{Q}_1} f_{\mathcal{D}_1} v_{\rm EW}$ and $m_s \sim \lambda_* f_{\mathcal{Q}_2} f_{\mathcal{D}_2} v_{\rm EW}$ are the masses of the up and strange quark, respectively, and the dimensionless $\lambda_*$ measures the typical size of the 5D Yukawa couplings in units of $k^{-1}$. For the case of constant fermion mass terms, the resulting limit on the mass of the first gluon KK mode is $m^{(1)}_\mathcal{G} \gtrsim (3/\lambda_*) (22 \pm 6)\, \text{TeV}$ \cite{Csaki:2008zd} (the error arises from the uncertainty in the down and strange quark masses). Using that the masses of the gluon KK modes are determined by $J_0(m^{(n)}_\mathcal{G}/m_\ir)\simeq 0$, this translates to the limit $m_\ir \gtrsim (3/\lambda_*)(9  \pm 3) \, \text{TeV}$ on the IR scale. 

Such a bound on $m_{\ir}$ introduces a little hierarchy problem. 
Numerous solutions have been proposed to solve it. Most of them introduce a new (partially gauged) flavour symmetry in the bulk, e.g. \cite{Santiago:2008vq,Csaki:2009wc,Redi:2011zi} (see \cite{Agashe:2013kxa,
vonGersdorff:2013rwa} for reviews of the current status). In all these works, the bulk mass of the fermions is assumed to be dominated by a constant.\footnote{The potential flavon-radion interplay and the possibility to use the Goldberger-Wise field as a flavon were discussed in ref.~\cite{Eshel:2011wz}, however in a context where extra flavour symmetries in the bulk remain the  key features and  the effect on fermionic profiles was not alluded to.} In contrast, the
bulk mass in model II is given by the VEV of the Goldberger-Wise field, and this affects the fermionic profiles.
Since the fermion wavefunctions are then suppressed in a large part of the bulk compared to the case of constant bulk masses (cf.~fig.~\ref{fig:wavefunctionsandcloc}),  the overlap integral with the gluon wavefunction is decreased. Accordingly, we expect $C^4_K$ to be smaller and the constraint on the KK gluon mass to be weakened. Indeed, matching the parameters $\rho$ for the relevant flavours to the benchmark point from ref.~\cite{Casagrande:2008hr} (cf.~sec.~\ref{sec:ModelI}), we find
\be 
\frac{\left[ C^4_K \right]_{c^{\rm loc}}}{\left[ C^4_K \right]_{c=\text{const.}}} \, \sim \, \frac{\left[(\gamma_{_{\mathcal{Q}_2}} \hspace{-.1cm}  + \gamma_{_{\mathcal{Q}_3}}) \, (\gamma_{_{\mathcal{D}_2}} \hspace{-.1cm}  + \gamma_{_{\mathcal{D}_3}})\right]_{c^{\rm loc}} }{\left[(\gamma_{_{\mathcal{Q}_2}} \hspace{-.1cm}  + \gamma_{_{\mathcal{Q}_3}}) \, (\gamma_{_{\mathcal{D}_2}} \hspace{-.1cm}  + \gamma_{_{\mathcal{D}_3}})\right]_{c=\text{const.}}} \, \sim \, \frac{1}{10} \, .
\ee
The limit then becomes $m^{(1)}_\mathcal{G} \gtrsim (3/\lambda_*)\,(7 \, \pm \, 2) \, \text{TeV}$ or $m_\ir \gtrsim (3/\lambda_*)(3  \pm 1) \, \text{TeV}$ in model II. For $\lambda_* \sim 3$, this is in the same ballpark as the constraint $m_\ir \gtrsim 1.9 \, \text{TeV}$ that arises from electroweak precision tests if a custodial symmetry is assumed \cite{Malm:2013jia,Bauer:2016lbe}. The modified fermion wavefunctions thus mitigate the RS-$CP$-problem 
that stems from $CP$-violation in $K$-$\overline{K}$-mixing in a very minimal way. This clearly deserves further investigation given the potentially important implications for the little hierarchy problem in RS.
Let us emphasize that this effect of coupling suppression between SM fermions and KK gluons works as long as 
the Higgs lives very near the IR brane.
If the Higgs is delocalized into the bulk and as delocalised as the KK gluon, on the other hand, there will be no effect.
The effect of weakening the bound on the KK scale from $\epsilon_K$ is thus maximal  when the Higgs lives exactly on the IR brane. 

\subsection{Constraints from one-loop contributions to the neutron EDM}
\label{neutronEDM}

Another important constraint arises from the neutron EDM. The most important, new contribution to this quantity arises at one-loop and is mediated by fermionic KK modes and the Higgs or the longitudinal components of the $Z$ \cite{Agashe:2004cp}. The corresponding Feynman diagram is shown in fig.~\ref{fig:FeynmanDiagram}(b). This gives rise to the effective operator
\be 
\mathcal{L} \, \supset \, -C_{d_n} \, \overline{d}_L  \sigma^{\mu \nu} d_R \, F_{\mu \nu}\, ,
\ee
where again $d_{L,R}$ is the down quark. We can estimate the Wilson coefficient as
\be
C_{d_n} \, \sim \, \frac{\lambda_*^2 \, e}{16 \, \pi^2} \frac{m_d}{\bigl[m^{(1)}_{\psi}\bigr]^2} \, ,
\ee
where $m_d \sim \lambda_* f_{\mathcal{Q}_1} f_{\mathcal{D}_1} v_{\rm EW}$ is the mass of the down quark and $m^{(1)}_{\psi}$ denotes the mass scale of the (lowest lying) fermionic KK modes in the loop. The EDM is proportional to the imaginary part of this Wilson coefficient. It can be shown that this imaginary part is unsuppressed and cannot be removed by field redefinitions, so that $\mbox{Im} \, C_{d_n} \sim |C_{d_n}|$ \cite{Agashe:2004cp}. The bound $d_n \leq 3 \cdot 10^{-26} \, e \, {\rm cm}$ \cite{Afach:2015sja} on the neutron EDM then translates to $m^{(1)}_{\psi} \gtrsim (\lambda_*/3) \cdot 26 \, {\rm TeV}$. For a fermion with constant bulk mass $c k$, the KK spectrum is determined by $J_{|c-1/2|}(m^{(n)}_{\psi}/m_\ir)\simeq 0$ \cite{Gherghetta:2000qt}. For $c \sim 1/2$, this gives the limit $m_\ir \gtrsim  (\lambda_*/3)\, 11 \, \text{TeV}$ on the IR scale.

The above estimates are modified in model II since the masses of the fermionic KK modes in the loop and their wavefunction overlaps with the IR brane (which are relevant for the vertices involving the Higgs) differ from the case of constant bulk masses. 
We have numerically determined these quantities for the first fermionic KK modes using the $\rho$-values that correspond to the benchmark point from ref.~\cite{Csaki:2008zd}. In table \ref{KKMassesTable}, we list the masses for the case of $y$-dependent and constant bulk masses. As one can see, the former are $(75 - 80)\%$ heavier than the latter (except for the left-handed top-bottom doublet and the right-handed top for which the mass increase is smaller). This can be understood as follows: The local bulk-mass parameter in eq.~\eqref{localcmod} is approximately constant near the IR brane, $c^{\rm loc} \approx -\beta \rho /k$. Since the light KK modes are localized in that region, we expect that their masses depend similarly on $c^{\rm loc}$ as for the case of constant bulk masses. The mass quantization condition for the latter case given above leads to masses which grow with $|c-1/2|$.
For $\beta =1.5 k^{3/2}$ and the $\rho$-values that correspond to the benchmark point from ref.~\cite{Csaki:2008zd}, $c^{\rm loc}$ near the IR brane is in the range $-1.7 \, ... \, -2.8$. This is thus much larger than the corresponding $c$-values for the case of constant bulk masses for the fermions (cf.~eq.~\eqref{BenchmarkPointParameters}) and leads to larger masses for the KK modes. We find that the wavefunction overlaps with the IR brane, on the other hand, change only by $(1-2)\%$. 
Since the contribution to the neutron EDM scales like $\bigl[m^{(1)}_{\psi}\bigr]^{-2}$, we expect that the limit on the IR scale $m_\ir$
is reduced in model II. However, to quantify this requires a more detailed study of the relative importance of the different fermionic KK modes in the loop. We leave this for future work. Nevertheless, it is clear that the modified fermion wavefunctions in model II also ease the RS-$CP$-problem that stems from the neutron EDM.

\begin{figure}
\centering
\begin{tabular}{|c|c|c|c|}
\hline
 & $m^{(1)}_{\mathcal{Q}_i}/m_\ir$ & $m^{(1)}_{\mathcal{U}_i}/m_\ir$ & $m^{(1)}_{\mathcal{D}_i}/m_\ir$ \\
 \hline
i=1 &  4.56 \, (2.54)  &   4.91 \, (2.81)  &   4.84 \, (2.76)\\
 \hline
i=2 & 4.44 \, (2.46)  &  4.52 \, (2.51) & 4.74 \, (2.68) \\
 \hline
i=3 &  4.34 \, (2.74) &   4.28 \, (3.61)  &  4.51 \, (2.50) \\
 \hline
\end{tabular}
\caption{Masses of the first fermionic KK modes in model II using the $\rho$-values that correspond to the benchmark point from ref.~\cite{Csaki:2008zd} and, in brackets, for the case of constant bulk masses.\label{KKMassesTable}}
\end{figure}

\subsection{Constraints from processes mediated by the Goldberger-Wise scalar}
\label{subsec:GWconstraints}

Next we consider processes mediated by the Goldberger-Wise scalar. We focus on the case that the radion is parametrically lighter than the KK modes of the Goldberger-Wise scalar. Then the mixing between the former and the latter can be neglected \cite{Chacko:2014pqa}. Similarly, modifications of the radion couplings due to the new couplings of the Goldberger-Wise scalar to the bulk fermions are suppressed \cite{Chacko:2014pqa}. Flavour- and $CP$-violating processes mediated by the radion are thus as usual (see e.g.~\cite{Azatov:2008vm,Huitu:2011zh}).

Let us first consider model II. We expand the Goldberger-Wise scalar around its VEV, $\phi=\langle \phi \rangle + \delta \phi$, and decompose it as $\delta \phi = \sqrt{k} \sum_n\phi^{(n)} f_\phi^{(n)}$ (see the appendix \ref{sec:app} for more details). We again work in the basis in which the Yukawa coupling in eq.~\eqref{BulkYukawaCoupling} for the bulk fermions is diagonal in flavour space. The coupling among the lightest KK modes of the Goldberger-Wise scalar and the fermions is then given by
\be 
\label{ActionGWcoupling}
S \, \supset \, - \int d^4 x \, \phi^{(1)} \left(\overline{\mathcal{Q}}_R^{(1)} \, \tilde{\rho}_{\mathcal{Q}} \, \mathcal{Q}_L^{(0)} \, + \, \text{h.c.}\right) \, ,
\ee
where $\tilde{\rho}_{\mathcal{Q}}$ is a diagonal matrix in flavour space with elements
\be 
\label{GWcoupling}
\left[\tilde{\rho}_{\mathcal{Q}}\right]_{ij} \, = \, \delta_{ij}\,  \rho_{\mathcal{Q}_i} \,  k^{3/2}  \int_{0}^{y_\ir} dy \, f_\phi^{(1)} \, f_{\mathcal{Q}_i R}^{(1)} \, f_{\mathcal{Q}_i L}^{(0)} \, .
\ee
Similar couplings exist for the right-handed quarks. After electroweak symmetry breaking, we rotate the quarks $\mathcal{U}^{(0)}_R \rightarrow U_{_{Ru}} u_R$ etc.~in order to obtain diagonal mass matrices. The coupling in eq.~\eqref{ActionGWcoupling} then induces flavour- and $CP$-violating processes. In particular, the Goldberger-Wise scalar contributes to the neutron EDM via processes of the type shown in fig.~\ref{fig:FeynmanDiagram}(b), where it replaces the Higgs. Since the KK modes are localized in the IR whereas the down quark lives towards the UV brane, the overlap integral in eq.~\eqref{GWcoupling} leads to a suppression factor of order $f_{\mathcal{Q}_1}f_{\mathcal{D}_1}$ in the amplitude similar to the process mediated by the Higgs. 
In addition we have some freedom in choosing the size of the couplings $\rho_{\mathcal{Q}_i}$ etc.~and the KK modes of the Goldberger-Wise scalar can be relatively heavy which can further suppress the contribution of the Goldberger-Wise scalar to the neutron EDM.
We therefore expect that the latter can be subdominant compared to the contribution mediated by the Higgs. However, we leave a more detailed study to future work.

Let us next consider model I. In this case the fermion wavefunctions are not modified but the coupling of the Goldberger-Wise scalar to the Yukawa operator on the IR brane results in new flavour- and $CP$-violating processes. Using the expression involving down-type quarks which corresponds to eq.~\eqref{NewYukawaCoupling} for the up-type quarks, we find for the coupling of the first Goldberger-Wise mode to the fermionic zero-modes and the Higgs
\be 
S\, \supset \, \int d^4 x \,  \frac{\phi^{(1)}}{m_\ir} \, H \, \overline{\mathcal{Q}}^{(0)}_L \, \tilde{\kappa}_{d} \,\mathcal{D}^{(0)}_R \, + \, \text{h.c.} \, ,
\ee
where $\tilde{\kappa}_{d}$ is a matrix in flavour space with elements
\be 
\left[\tilde{\kappa}_{d}\right]_{ij} \, = \, f_{\mathcal{Q}_i} \, f_{\mathcal{D}_j} \left(\kappa_d\right)_{ij} k^{7/2} B \, .
\ee
As before, $f_{\mathcal{Q}_i} \equiv e^{k y_\ir/2} f^{(0)}_{{\mathcal{Q}_i L}}(y_\ir)$ etc.~are the wavefunction overlaps of the zero-modes with the IR brane and 
\be 
B \, \equiv \, \frac{\partial_y f^{(1)}_\phi\bigr|_{y_\ir}}{k e^{k\,  y_\ir}} \, = \, \frac{- 2 \, b_\ir\,\mathcal{N}^{(1)}_\phi \,  e^{k \, y_\ir}}{\pi \, m_\phi^{(1)} Y_{1+\epsilon}\left(m_\phi^{(1)}/m_\ir \right)/m_\ir \, + \,\pi \, (b_\ir - \epsilon) \, Y_{2+\epsilon}\left(m_\phi^{(1)}/m_\ir \right)} \, .
\ee
The normalization constant $\mathcal{N}^{(1)}_\phi$ is given in eq.~\eqref{Def_N} and $b_\ir$ is defined in eq.~\eqref{Def_b}. For example for $\epsilon=1/20$ and $b_\ir=10$, we find $B =-5.32$.   
Rotating the quarks $\mathcal{D}^{(0)}_R \rightarrow U_{_{Rd}} d_R$ etc.~after electroweak symmetry breaking in order to obtain diagonal mass matrices, this gives
\be 
S\, \supset \, \int d^4 x \, \phi^{(1)} \, \overline{d}_L \, \hat{\kappa}_d \,  d_R \, + \, \text{h.c.} \, ,
\ee
where the elements of the matrix $\hat{\kappa}_d$ are
\be 
\left[\hat{\kappa}_d\right]_{ij} \, \sim \,f_{\mathcal{Q}_i} \, f_{\mathcal{D}_j} \, \frac{v_{_{\rm EW}}}{m_\ir}  \, \left(\kappa_d\right)_{ij} k^{7/2} B \,\sim \,f_{\mathcal{Q}_i} \, f_{\mathcal{D}_j} \,B\, \frac{v_{_{\rm EW}}}{m_\ir}\, \frac{\ell_5^{1/3}}{\ell_4^{1/2}}\left(\frac{k}{M_5}\right)^{7/2} \, .
\ee
In the last step, we have used the estimate \eqref{NDAestimate} from naive dimensional analysis. Notice that the unitary rotation matrices $U_{_{Rd}}$ etc.~do not change the dependence on $f_{\mathcal{Q}_i} f_{\mathcal{D}_j}$ in $\hat{\kappa}_d$ versus $\tilde{\kappa}_d$. Due to this coupling, the Goldberger-Wise scalar can in particular contribute to $\epsilon_K$ via tree-level exchange similar to the gluon in fig.~\ref{fig:FeynmanDiagram}(a). Integrating out the first KK mode of the Goldberger-Wise scalar, we can estimate the Wilson coefficient of the operator in eq.~\eqref{EffectiveOperator} as
\be 
C^4_K \, \sim \,  \frac{m_d \, m_s}{m_\ir^2 \, \lambda_*^2}\, \frac{B^2}{\bigl[m^{(1)}_\phi\bigr]^2} \, \frac{\ell_5^{2/3}}{\ell_4} \left(\frac{k}{M_5}\right)^7 \, .
\ee
We see that this is suppressed compared to the Wilson coefficient in eq.~\eqref{GluonMediatedC4K} that arises from gluon exchange by factors $(v_{\rm EW}/m_\ir)^2$ and $(k/M_5)^7$. 
We therefore again expect that constraints on flavour- and $CP$-violation due to the Goldberger-Wise scalar can be readily fulfilled but leave a more detailed study to future work.

\section{Interpretation of the models in the dual CFT}
\label{sec:CFTint}

The Randall-Sundrum model has a dual description in terms of a strongly-coupled CFT via the AdS/CFT correspondence \cite{Maldacena:1997re}. The presence of the UV brane corresponds to the CFT being coupled to gravity \cite{Gubser:1999vj} while the IR brane is dual to the spontaneous breaking of conformal invariance in the IR \cite{ArkaniHamed:2000ds,Rattazzi:2000hs}. 

Stabilizing the extra dimension by the Goldberger-Wise mechanism is dual to deforming the CFT at the cutoff scale $\Lambda_\uv \sim k$ by an almost marginal operator $\mathcal{O}_\phi$ of dimension $ 4 + \epsilon$,
\be 
\mathcal{L} \, = \, \mathcal{L}_{\rm CFT }\, + \, \frac{B}{ \Lambda_\uv^{3/2+\epsilon}} \mathcal{O}_\phi\, ,
\ee
where $\epsilon$ and $B$ are the parameters that determine the VEV of the Goldberger-Wise scalar in eq.~\eqref{vevprofile}. This operator runs slowly when going towards lower energies until it eventually triggers the breaking of conformal invariance at a scale 
\be
{\Lambda}^{\rm min}_\ir \sim k \, \sigma^{\rm min}_\ir.
\ee
Moving the radion VEV away from its value $\sigma_\ir^{\rm min}$ at the minimum of the Goldberger-Wise potential to some value $\sigma_\ir$ then corresponds to changing the confinement scale of the theory\footnote{The groundstate of this theory differs from that for a confinement scale ${\Lambda}^{\rm min}_\ir$ and is obtained by minimizing the energy $\langle \rho | H_{\rm CFT}| \rho \rangle$ over all states $|\rho \rangle$ that keep $\langle \mathcal{O}_\phi \rangle= \langle \rho |\mathcal{O}_\phi| \rho \rangle$ fixed at the value given in eq.~\eqref{OperatorVEV} below \cite{ArkaniHamed:2000ds}. } from ${\Lambda}^{\rm min}_\ir$ to $\Lambda_\ir \sim k \, \sigma_\ir$. Furthermore, the parameter $A$ in eq.~\eqref{vevprofile} is dual to the VEV of the operator,
\be 
\label{Arelation}
\langle\mathcal{O}_\phi \rangle \, = \, \Lambda_\uv^{5/2+\epsilon} \, A \, .
\ee

A fermion with a constant mass term $c k > -k/2$ in the bulk of a Randall-Sundrum model is dual to the system \cite{Contino:2004vy}
\be 
\label{DualLagrangian}
\mathcal{L} \, \supset \, \mathcal{L}_{\rm CFT} \, + \, i \, \mathcal{Z}  \, \bar{\psi}_L \gamma^\mu \partial_\mu \psi_L \, + \, \frac{\omega}{\Lambda_\uv^{\Delta-5/2}} \left( \bar{\psi}_L \mathcal{O}_R \, + \, \text{h.c.} \right) \, ,
\ee
where $\psi_L$ is a left-handed, massless fermion, $\mathcal{O}_R$ is a fermionic CFT operator with dimension 
\be
\Delta = 3/2+|c+1/2|
\ee
 and $\mathcal{Z}$ and $\omega$ are dimensionless constants.\footnote{An alternative description involves a right-handed instead of the left-handed fermion \cite{Contino:2004vy}.} 
 Let us focus on a bulk fermion with boundary conditions leading to a left-handed massless zero-mode. According to the dictionary from ref.~\cite{Contino:2004vy}, the dual theory in this case has no massless composite states once conformal invariance is broken. The spectrum therefore contains exactly one massless fermion which generically is an admixture of $\psi_L$ with the composite states generated by the operator $\mathcal{O}_R$. This state is dual to the zero-mode of the bulk fermion. If $\Delta>5/2$, the operator in eq.~\eqref{DualLagrangian} which mixes $\psi_L$ and the composite states is irrelevant and the massless state therefore consists dominantly of $\psi_L$. In the opposite case $\Delta<5/2$, the mixing operator is relevant and the massless state has a significant composite contribution. On the Randall-Sundrum side, this corresponds to $c>1/2$ and a UV-localized zero-mode and $c<1/2$ and an IR-localized zero-mode, respectively.

In model II, the bulk fermions instead have position-dependent masses $k c^{\rm loc}(y)$. Since the position along the extra dimension corresponds to the RG scale of the dual theory, $e^{- k y} \Leftrightarrow \mu/\Lambda_\uv$, we expect that the dual description is again given by eq.~\eqref{DualLagrangian} but with a large anomalous dimension 
\be 
\Delta(\mu) \, = \, \frac{3}{2} \,+ \, \left|c^{\rm loc}\Bigl(\frac{1}{k}\log \frac{\Lambda_\uv}{\mu}\Bigr) \, + \, \frac{1}{2} \right| \, .
\ee
We will now show that this reproduces the Yukawa couplings that we have found in the 5D description. To this end, we will focus on the simple profile for the Goldberger-Wise scalar in eq.~\eqref{simplescalarvev} but the derivation can be extended to the other profiles considered in this paper too. Using eq.~\eqref{localc}, the anomalous dimension then reads
\be
\label{AnomalousDimensionSP}
\Delta(\mu) = 2+ \tilde{c}  \left(\frac{\mu }{ \Lambda_\uv}\right)^\epsilon 
\ee
for $\tilde{c}>-1/2$. We define the dimensionless parameter 
\be 
\xi(\mu) \, \equiv \, \frac{\omega(\mu)}{\sqrt{\mathcal{Z}(\mu)}} \Bigl(\frac{\mu}{\Lambda_\uv}\Bigr)^{\Delta(\mu)-5/2} 
\ee
which measures the mixing between $\psi_L$ and the CFT (or the composite states once conformal invariance is broken). It satisfies the RG equation \cite{Contino:2004vy}
\be 
\label{RGequation}
\mu \frac{d \xi}{d \mu} \, = \, \Bigl( \Delta \, - \, \frac{5}{2} \Bigr) \, \xi \, + \, \frac{\eta \, N}{16 \, \pi^2} \, \xi^3 \, ,
\ee
where $N$ is the number of colors of the CFT, $\eta = \mathcal{O}(1)$ and the second term arises from the CFT contribution to the wavefunction renormalization  $\mathcal{Z}$ of $\psi_L$.

First we consider the case that $\Delta > 5/2$ at the cutoff scale $\Lambda_\uv$ (corresponding to $\tilde{c}>1/2$).
The first term in the RG equation then reduces the coupling when going to lower energies and both terms become comparable at some scale $\tilde{\mu}$. The mixing parameter at that scale is 
\be
\xi(\tilde{\mu})\approx 4 \pi \, \sqrt{(\Delta(\tilde{\mu})-5/2)/(\eta N)} 
\ee
and we expect that $\tilde{\mu}\approx \Lambda_\uv$. Assuming that $\Delta > 5/2$ over a sufficiently large range of energies, we can neglect the second term over the remaining RG evolution and integrate the RG equation in closed form. At the scale $\Lambda_\ir$ this gives:
\be 
\label{mixing}
\xi(\Lambda_\ir) \, \approx \, 4 \pi \sqrt{\frac{\tilde{c}-\frac{1}{2}}{\eta N}} \, \sqrt{\frac{\Lambda_\uv}{\Lambda_\ir}} \, e^{-\frac{\tilde{c}}{\epsilon}} \, e^{\frac{\tilde{c}}{\epsilon}\bigl(\frac{\Lambda_\ir}{\Lambda_\uv}\bigr)^\epsilon} \, .
\ee
The above approximations are in particular valid for the case of small mixing, $\xi(\Lambda_\ir) \ll 1$. 

Next we consider the opposite case of strong mixing $\xi(\Lambda_\ir) \gtrsim 1$. This can occur if $\Delta <5/2$ over a sufficiently large range of energies during the RG evolution. Then the second term in the RG equation can no longer be neglected. Assuming that $|\tilde{c} (\mu / \Lambda_\uv)^\epsilon|  \ll 1/2$ at energies somewhat above $\Lambda_\ir$ so that $\Delta(\mu) -5/2 \approx -1/2$, we can again integrate the RG equation in closed form. We then find that the mixing parameter runs to the fixed point
\be 
\label{FixedPoint}
\xi \, = \, \sqrt{\frac{8 \pi^2}{N  \eta}} \, .
\ee

\begin{figure}[t]
\centering
\includegraphics[width=6cm]{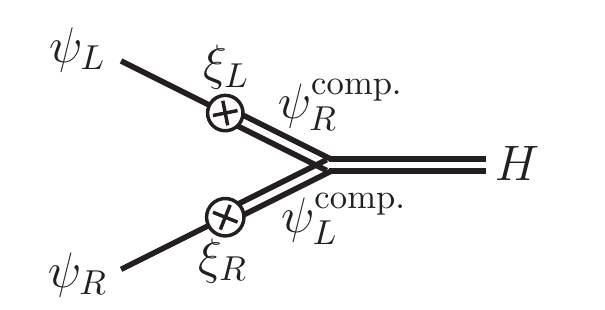}
\caption{\label{fig:YukawaCoupling} \small Origin of the Yukawa couplings to SM fermions in the dual CFT. }
\end{figure}

So far we have only discussed the dual description of bulk fermions with left-handed zero-modes. Similarly, bulk fermions with right-handed zero-modes are described by eq.~\eqref{DualLagrangian} with a right-handed, massless fermion $\psi_R$ which mixes with an operator $\mathcal{O}_L$. We identify the massless states which arise from the combined Lagrangian with the left- and right-handed fields of the SM. Each has its own mixing parameter $\xi_L$ or $\xi_R$. 
The Higgs on the IR brane is dual to a composite state and will generically have large couplings to other composite states. The size of the Yukawa couplings to SM fermions is then controlled by the degree of compositeness of the massless states in the dual theory and thus by the mixing parameters $\xi_L$ and $\xi_R$. In particular for $\xi_L,\xi_R \ll 1$, the massless states consist dominantly of $\psi_L$ and $\psi_R$ and the Yukawa couplings are suppressed by the small mixing parameters:
\be
y(\Lambda_{\ir}) \, \propto \, \xi_L(\Lambda_\ir) \times \xi_R(\Lambda_\ir) \, .
\ee 
The corresponding Feynman diagram is shown in fig.~\ref{fig:YukawaCoupling}. 
Assuming that the dual theory is a gauge theory with large number of colors $N$ (as is implied by full string-theory examples of the AdS/CFT correspondence), we can determine the prefactor in the above relation. In this case, the overlap between an operator $\mathcal{O}$ and composite fermions $\psi_{\rm comp.}$ is given by $\langle 0 | \mathcal{O} \, \psi_{\rm comp.}|0\rangle \sim \sqrt{N}/4 \pi$ \cite{Witten:1979kh}. Furthermore, the vertex between three composite states is $\Gamma_3 \sim 4 \pi /\sqrt{N}$ \cite{Witten:1979kh}. Using eq.~\eqref{mixing} for the left- and right-handed state, the resulting Yukawa couplings are
\be 
\label{DualModifiedYukawaCouplings}
y(\Lambda_\ir) \, \approx \, \sqrt{\tilde{c}_{_L}-\frac{1}{2}} \sqrt{\tilde{c}_{_R}-\frac{1}{2}} \frac{4 \pi}{\sqrt{N }\eta}\frac{\Lambda_\uv}{\Lambda_\ir} \,  e^{-\frac{\tilde{c}_{_L}+\tilde{c}_{_R}}{\epsilon}\bigl( 1-\bigl(\frac{\Lambda_\ir}{\Lambda_\uv}\bigr)^\epsilon \bigr)}  \, .
\ee
The limit of small mixing, $\xi_L, \xi_R \ll 1$, 
corresponds to fermions which are localized towards the UV brane. The Yukawa coupling from the 5D description is then well approximated by eq.~\eqref{YukawaUVapprox}.
Identifying $\eta = 1/2$ and $\lambda k = 4 \pi/\sqrt{N}$, where $\lambda$ is the 5D Yukawa coupling, we see that eq.~\eqref{DualModifiedYukawaCouplings} reproduces the Yukawa coupling from the 5D description. 


Similarly in the case of strong mixing, using eq.~\eqref{FixedPoint} for the left- and right-handed state gives
\be 
\label{DualModifiedYukawaCouplings2}
y(\Lambda_\ir) \, \approx \, \frac{1}{2 \eta} \frac{4 \pi}{\sqrt{N}} \, .
\ee
The case of strong mixing corresponds to fermions which are localized towards the IR brane. The Yukawa coupling from the 5D description is then well approximated by eq.~\eqref{YukawaIRapprox}. Again identifying $\eta = 1/2$ and $\lambda k = 4 \pi/\sqrt{N}$, we see that eq.~\eqref{DualModifiedYukawaCouplings2} reproduces the Yukawa coupling from the 5D description.

We found that the coupling of KK gluons to SM fermions can be reduced when taking the fermionic bulk masses to be $y$-dependent (while keeping the 4D Yukawa couplings of the SM fermions fixed).
In the CFT language, this means that the coupling of SM fermions to composite gluons is reduced when 
changing the scaling dimensions of the fermionic operators while keeping the amount of compositeness of the SM fermions fixed.

We can also apply the AdS/CFT dictionary to model I. Using eqs.~\eqref{leadingA}, \eqref{hierarchyrelation} and \eqref{Arelation}, we see that 
moving the radion VEV away from the minimum of the Goldberger-Wise potential changes the VEV of the operator,
\be 
\label{OperatorVEV}
 \langle\mathcal{O}_\phi \rangle \, \sim \, \Lambda_\ir^{4+2\epsilon} \frac{B}{\Lambda_\uv^{3/2+\epsilon}} \left(\left(\frac{{\Lambda}^{\rm min}_\ir}{\Lambda_\ir}\right)^\epsilon \left(1-\sqrt{\frac{\epsilon}{4}} \right)-1 \right) \, .
\ee
When $\Lambda_\ir=\Lambda_\ir^{\rm min}$, this VEV is suppressed as $\langle \mathcal{O}_\phi \rangle \propto \sqrt{\epsilon}$. It increases when moving away from the minimum.
The new contribution from the derivative coupling in eq.~\eqref{NewYukawaCoupling} to the Yukawa coupling for the top quark can be written as
\be 
\label{eq:CFTmodel1}
[\delta y_u]_{33} \, \sim \,  \left. \frac{\partial_y \langle \phi \rangle}{k^{5/2}}\right|_{y=y_{\rm IR}}  \sim \, \frac{4\, \langle\mathcal{O}_\phi \rangle}{\Lambda_\ir^{4+\epsilon}} \, - \,   \frac{\epsilon \, B \, \Lambda_\ir^\epsilon}{\Lambda_\uv^{3/2+\epsilon}} \, ,
\ee
where we have used eqs.~\eqref{vevprofile} and \eqref{Arelation} and set $\kappa_u \sim k^{-7/2}$. Note that for the  UV-localized flavours, the Yukawa couplings are suppressed compared to eq.~(\ref{eq:CFTmodel1}) due to the small overlap of their wavefunctions with the IR brane (cf.~eq.~\eqref{4dYukawa}). This is dual to small mixing $\xi_L,\xi_R \ll 1$ between the fundamental fermions and the composite states of the broken CFT as discussed above.
We thus see that in the dual description, the top Yukawa coupling gets a contribution proportional to $\langle \mathcal{O}_\phi \rangle$ via the first term in eq.~(\ref{eq:CFTmodel1}).
Notice also that at the scale $\Lambda_\ir^{\rm min}$ the second term is suppressed by $\sqrt{\epsilon/4}$ relative to the first term and becomes even less important as $\Lambda_\ir$ decreases.
The change in the Yukawa coupling when the IR brane is moved then dominantly arises from the change in $\langle \mathcal{O}_\phi \rangle$ when the dual broken CFT is in states with different confinement scales $\Lambda_\ir$.    
In this description, it is also clear that in model I (in contrast with model II) we cannot get contributions $\delta y$ of order one for species other than the top quark since at initial times, $\langle\mathcal{O}_\phi \rangle$ is small and then evolves to values of order $\Lambda_\ir^{4+\epsilon}$.

\section{Conclusions}
\label{sec:conclusion}

We have shown how the Randall-Sundrum model with Goldberger-Wise stabilisation offers a natural display of the cosmological  emergence of the flavour structure in the standard model. Our main new results are contained in secs.~\ref{sec:ModelI}, \ref{sec:ModelII} and \ref{sec:flavourconstraints}.
In particular, we have shown how coupling the Goldberger-Wise scalar to the standard model fermions on the IR brane or in the bulk can lead to an effective 4D Yukawa coupling which increases across the bubble walls during the electroweak phase transition. 
This then provides a new source of $CP$-violation which allows for electroweak baryogenesis from the CKM matrix, and may also be relevant for cold baryogenesis.
It will be interesting to study this mechanism further and to understand  whether certain 4D flavour models could fall into this category. In particular, because Randall-Sundrum models are holographic duals of 4D strongly coupled theories, our findings may be useful for the investigation of flavour cosmology in composite Higgs models.
 
We now compare our findings with the results of another analysis of Yukawa variation during the electroweak phase transition, in Froggatt-Nielsen models~\cite{Baldes:2016gaf}.
In this  context, it was found that a very light flavon (i.e.~below the electroweak scale) is necessary in order to affect the values of the Yukawa couplings during the eletroweak phase transition. 
The main reason for this is the assumed structure of the polynomial two-field (Higgs and flavon) scalar potential. 
Such a light flavon, however, is in tension with experimental constraints. 
In Randall-Sundrum models, both the dependence of the Yukawa couplings on the radion and
the interplay between Higgs and radion are of a very different nature. It is possible to have a large variation of the Yukawa couplings during the electroweak phase transition for radion masses around or above the electroweak scale.
The key point is that the Higgs mass parameter is controlled by the radion VEV while in the Froggatt-Nielsen implementation of ref.~\cite{Baldes:2016gaf}, it is a constant like in the standard model.
We can therefore expect to build successful models of Yukawa coupling variation during the electroweak phase transition in models where the Higgs mass parameter is dynamical as well and controlled by parametrically slightly heavier, $\mathcal{O}$(TeV) scale new physics, similar to what happens in the Randall-Sundrum construction. 

Finally and interestingly, in our construction in which the 5D fermionic mass terms  are not constant but result from the coupling to the Goldberger-Wise scalar,  the fermionic profiles are suppressed in much of the bulk compared to the case of constant mass terms. This suppresses their overlap with KK modes, and thereby weakens the constraints from $CP$-violating processes in Randall-Sundrum constructions.

\section*{Acknowledgments}
We thank Kaustubh Agashe, Iason Baldes, Sebastian Bruggisser, Tony Gherghetta, Thomas Konstandin and Gilad Perez for useful discussions.

\appendix

\section{Appendix: KK expansion of the Goldberger-Wise scalar}
\label{sec:app}

Our discussion applies to the Goldberger-Wise scalar with both the potentials considered in the original paper and discussed in sec.~\ref{sec:ReviewGW} and the modified potentials considered in sec.~\ref{sec:modifiedGW}. 
We assume that the radion is parametrically lighter than the IR scale. Then the mixing between the radion and the KK modes of the Goldberger-Wise scalar is suppressed by the ratio of their masses and can be neglected to a good approximation (see appendix A in \cite{Chacko:2014pqa}). We expand the Goldberger-Wise scalar around its VEV, $\phi = \langle \phi \rangle + \delta \phi$, and decompose it as $\delta \phi = \sqrt{k} \sum_n\phi^{(n)} f_\phi^{(n)}$. The bulk equation of motion and the boundary conditions read
\begin{gather}
\left( \partial_y^2   \, - \, 4 k \, \partial_y \, - \, m_\phi^2  \, + \, e^{2 k y} (m^{(n)}_\phi)^2 \right)\, f_\phi^{(n)} \, = \, 0\\
\left( \partial_y - k \, b_\uv \right) f_\phi^{(n)}\bigr|_{y=0} \, = \, 0 \quad \qquad \left( \partial_y + k \, b_\ir \right) f_\phi^{(n)}\bigr|_{y=\pi R} \, = \, 0\, ,
\end{gather} 
where 
\be 
\label{Def_b}
b_{\uv,\ir} \equiv  \frac{1}{2 k}\frac{\partial^2 V_{\uv,\ir}}{\partial \phi^2}\bigr|_{\phi=\langle\phi \rangle} \, .
\ee
This is solved by
\be 
f_\phi^{(n)}(y) \, = \, \mathcal{N}^{(n)}_\phi  e^{2 ky} \left(J_{2+\epsilon}\bigl(m^{(n)}_\phi e^{ky}/k\bigr)\, + \, b_n\bigl(m^{(n)}_\phi/k,b_\uv \bigr) \, Y_{2+\epsilon}\bigl(m^{(n)}_\phi e^{ky}/k\bigr)\right)
\ee
with
\be 
b_n\left(r,b\right) \, \equiv \,- \frac{r \, J_{1+\epsilon}\left(r\right)\, - \,(b+\epsilon)J_{2+\epsilon}\left(r\right) }{r \, Y_{1+\epsilon}\left(r\right)\, - \,(b +\epsilon)Y_{2+\epsilon}\left(r\right)} \, .
\ee
Away from the UV brane this is well approximated by
\be 
\label{GWKKmodewavefunction}
f_\phi^{(n)}(y) \, \simeq \, \mathcal{N}^{(n)}_\phi e^{2 ky} J_{2+\epsilon}\bigl(m^{(n)}_\phi e^{ky}/k\bigr)  \, .
\ee
The normalization constant of the wavefunction is given by 
\be
\begin{split}
\label{Def_N}
\bigl( \mathcal{N}^{(n)}_\phi\bigr)^{-2} \, & = \, \int_0^{y_\ir} dy \, k \, e^{2 ky} \, \left(J_{2+\epsilon}\bigl(m^{(n)}_\phi e^{ky}/k\bigr) +  b_n\bigl(m^{(n)}_\phi/k,b_\uv \bigr) \, Y_{2+\epsilon}\bigl(m^{(n)}_\phi e^{ky}/k\bigr)\right)^2 \\
& \simeq \, \frac{1}{2}\, e^{2 k y_\ir}  \, \bigl(J_{2+\epsilon}\bigl(m^{(n)}_\phi /m_\ir\bigr) \bigr)^2\, \left[1\, + \, \frac{m_\ir^2}{(m^{(n)}_\phi)^2}\, \left( (4+2\, \epsilon)\, b_\ir\,  +  \, b_\ir^2 \right) \right] \, .
\end{split}
\ee
The mass quantization condition is given by $b_n\bigl(m^{(n)}_\phi/k,b_\uv \bigr) =b_n\bigl(m^{(n)}_\phi/m_\ir,-b_\ir \bigr) $. Expanding this for $\smash{m^{(n)}_\phi \ll k}$, the condition simplifies to
\be 
\frac{m^{(n)}_\phi}{m_\ir} \, J_{1+\epsilon}\bigl(m^{(n)}_\phi /m_\ir\bigr)\, + b_\ir\, J_{2+\epsilon}\bigl(m^{(n)}_\phi /m_\ir\bigr) \, \simeq \, 0 \, .
\ee


\end{document}